\documentclass[footinbib,twocolumn,showpacs,amsmath,amstex,amssymb,mathfonts,superscriptaddress,prx]{revtex4}
\usepackage{graphicx}
\usepackage{color}
\usepackage{bm}

\usepackage{graphicx}
\usepackage{color}
\usepackage{bm}
\usepackage{amsmath}
\usepackage{amssymb}
\usepackage{amsthm}
\usepackage{amsfonts}
\usepackage{multirow}
\usepackage{makecell}
\usepackage{float}
\usepackage{comment}
\usepackage{enumitem}
\usepackage{verbatim}
\usepackage{mathtools}

\begin{document}

\title{Composing parafermions: a construction of $Z_{N}$ fractional quantum Hall systems and a modern understanding of confinement and duality}
\author{Yoshiki Fukusumi} 
\affiliation{Division of Physics and Applied Physics, Nanyang Technological University, Singapore 637371.}

\date{\today}
\begin{abstract}

In this work, we propose a modern view of the integer spin simple currents which have played a central role in discrete torsion. We reintroduce them as nonanomalous composite particles constructed from $Z_{N}$ parafermionic field theories. These composite particles have an analogy with the Cooper pair in the Bardeen-Cooper-Schrieffer theory and can be interpreted as a typical example of anyon condensation. Based on these $Z_{N}$ anomaly free composite particles, we propose a systematic construction of the cylinder partition function of  $Z_{N}$ fractional quantum Hall effects (FQHEs). One can expect realizations of a class of general topological ordered systems by breaking the bulk-edge correspondence of the bosonic parts of these FQH models. We also give a brief overview of various phenomena in contemporary condensed matter physics, such as $SU(N)$ Haldane conjecture, general gapless and gapped topological order with respect to the quantum anomaly defined by charges of these simple currents and bulk and boundary renormalization group flow.
Moreover, we point out an analogy between these FQHEs and 2d quantum gravities coupled to matter, and propose a $Z_{N}$ generalization of supersymmetry known as "fractional supersymmetry" in the composite parafermionic theory and study its analogy with quark confinement. Our analysis gives a simple but general understanding of the contemporary physics of topological phases in the form of the partition functions derived from the operator formalism.

\end{abstract}

\maketitle 

\section{Introduction}

Recently, a general construction for fractional quantum Hall (FQH) states is proposed by Schoutens and Wen in \cite{Schoutens:2015uia}, by using the $Z_{N}$ simple currents or $Z_{N}$ parafermions in a typical case\cite{Fateev:1985mm}. Their construction is quite simple and universal, just by assuming the existence of the $Z_{N}$ simple currents and imposing single-valuedness of  the wavefunctions for the FQH systems. Hence, in principle, one can construct an arbitrary FQH wavefunction started from any conformal field theory (CFT) with $Z_{N}$ simple currents. Especially for $SU(N)_{K}$ Wess-Zumino-Witten (WZW) models, realizations of such $Z_{N}$ FQH states have been proposed in \cite{Fuji_2017} by using the coupled wire construction\cite{Kane_2002,Teo_2014,Sagi_2014,Imamura_2019}. Therefore, one can say that the existence of the bulk wavefunction of $Z_{N}$ FQHE has been established. However, only with their constructions, there still exist at least two difficulties to obtain the candidates of protected edge modes which can define a characteristic property of the topological phase of matter\cite{2013PhRvX...3b1009L}. The first is the difficulty to obtain multipoint correlation functions of CFT which corresponds to FQHE wavefunctions. Without some particular structure of the CFT, the calculation for the multipoint correlation functions needs analysis of higher-order ordinal differential equations or corresponding Dotsenko-Fateev integral\cite{Dotsenko:1984nm,Dotsenko:1984ad,Estienne:2009mp,Estienne:2010as,Estienne:2011qk}. The second is the difficulty to construct the braiding and modular property of CFTs which result in the Lagrangian subalgebra \cite{2013PhRvX...3b1009L} of the CFTs in the cylinder geometry. In this work, we propose a systematic way to overcome the second difficulty by considering the simple charges of the $Z_{N}$ simple currents in CFT which is a generalization of Lieb-Schultz-Mattis(LSM) type anomaly\cite{Kong:2014qka,Cho:2017fgz,Lanzetta:2022lze,Li:2022drc}. 

The simple charge of an operator in conformal field theory\cite{Fuchs:1997af,Schweigert:2000ix}, which corresponds to the symmetry charge of the corresponding operator in the lattice model, is one of the most important structures to study renormalization group (RG) flow under symmetry and gauging finite group symmetry\cite{Bhardwaj:2017xup,Tachikawa:2017gyf}. At first sight, these two operations, analyzing the RG flow and gauging symmetry seem different. However, as can be seen in the development of  conformal field theory and Haldane conjecture\cite{Haldane:1983ru,Haldane:1982rj,Cabra:1998vw,Lecheminant:2015iga,Lajko:2017wif,Wamer:2019oge,Yao:2020dqx}, these physical and mathematical structures should have some common underlying structures. For example, $SU(2)_{K}$ WZW model, where $K$ is the level of the model corresponding to the spin $K/2$ in the spin chain, is represented as the composition of $Z_{K}$ parafermion and $U(1)$ theory, $Z_{K}\times U(1)$. Hence by gapping the parafermion part, one can obtain $U(1)$ theory but the anomaly of $Z_{2}$ symmetry in the $SU(2)_{K}$ model becomes LSM type anomaly. In the lattice model, this results in the connectivity problem to $SU(2)_{1}$ point for higher spin $K/2$ Heisenberg XXZ chain and the structure $K \ (\text{mod.} 2)$ appear\cite{PhysRevB.46.2896,Alcaraz_1989,Cabra:1998vw}. In this case, only the model with $K=1 \ (\text{mod.}2)$ can be connected to $SU(2)_{1}$ point. On the other hand, the $SU(2)_{K}$ model has a $Z_{2}$ simple current, and this becomes nonanomalous for $K=0 \ (\text{mod}. 2)$ and one can obtain fermionic representation by gauging this $Z_{2}$ symmetry coming from $Z_{2}$ simple current in that case\cite{Hsieh:2020uwb,Fukusumi:2021zme}\footnote{It should be noted that there exists nonanomalous $Z_{2}$ symmetry generated by the charge conjugation in the $Z_{K}$ parafermion model. Historically, this nonanomalous symmetry appear as the $Z_{2}$ symmetry of $D_{K}=Z_{2}\times Z_{K}$ representation of the Fateev-Zamolodchikov parafermion model\cite{Fateev:1985mm}. More recently, it has typically appeared as (Majorana) fermionization of three state Potts model in \cite{Hsieh:2020uwb}. Because of anomaly matching of $SU(2)_{K}=Z_{K}\times U(1)$ model, the $Z_{2}$ anomaly of the model survives by gapping $Z_{K}$ parafermion. It should be noted that this charge conjugation can be outside of primary fields in the original $Z_{K}$ model. In that case, one cannot use the terminology "simple current" as the symmetry operator in a rigorous sense.}. Hence, these two concepts, RG flow and gauging finite group symmetry are related by the quantum anomaly and the fermionic representation of the model.

Roughly speaking, compared with an anomaly free case, an anomalous theory tends to become gapless because an anomaly of a symmetry restricts the permitted perturbative operators in Hamiltonian more strongly by the symmetry charge. In this sense, the gaplessness can be protected by symmetry in an anomalous theory. In an anomaly free case, because there exists less anomaly obstruction, the (mass) condensations occur and the system tends to become gapped. This is a simple understanding of the Haldane conjecture by considering $Z_{2}$ anomaly. Of course, this is a rough statement, and for justification in the specific models, the RG analysis with the table of the possible operators (possibly, both in bulk and boundary) in the theories is necessary. For an simplest example, one can consider that the spin $1/2$ anisotropic Heisenberg XXZ chain, which is $Z_{2}$ anomalous, becomes gapped under the strong anisotropy, whereas the higher spin integer anisotropic Heisenberg XXZ chain becomes gapless under the weak anisotropy\cite{PhysRevB.46.2896,Alcaraz_1989}.

More recently, the defect analog of the RG flow, called RG domain wall, has been proposed and applied to several important classes of models\cite{Gaiotto:2012np,Poghosyan:2014jia,Brunner:2015vva,Stanishkov:2016pvi,Poghosyan:2022mfw}. Putting this framework and anomaly analysis together, one can estimate the RG connectivity of the models by considering coupling the RG-connected systems and their simple charges of the product of simple currents in the original theories. Related proposals with an emphasis on the nontrivial conformal defects and analysis with the quantum impurity problems can be seen in \cite{2014NuPhB.885..266K,2015JHEP...07..072K,2007JHEP...04..095Q}. Moreover, this may give a unified understanding also for the boundary of $2+1$ dimensional topologically ordered systems. By stacking two topologically ordered systems and considering the simple charge of the product theory, one can obtain the equivalence relations, so-called Witt equivalent in the contemporary mathematical and theoretical condensed physics\cite{Kong:2022cpy}. For example, this aspect related to topological (nonprotected or condensed) boundary conditions has been discussed in \cite{Kaidi:2021gbs}. Hence, in other words, the Haldane conjecture and the classification problems of the protected edge modes of topologically ordered states can be understood in the language of the anomaly matching condition of the simple currents. Here, we note the importance to start the RG analysis of the connected systems with enough symmetries or central charges where the symmetry acts on site. 

 On the other hand, there exists another approach relating the Haldane conjecture and the structure of CFT, by investigating modular properties of the orbifolded model and symmetry action to BCFT\cite{Furuya:2015coa,Numasawa:2017crf,Kikuchi:2019ytf}. Their analysis is based on the modular $T$ and $S$ property of the systems under the symmetry action and can be understood as a classification based on the t'Hooft anomaly. Unfortunately, without some modifications, these approaches cannot produce the $K (\text{mod}.N)$ periodicity of $SU(N)_{K}$ systems \cite{Yao:2018kel} which is expected from the $Z_{N}$ anomalies from the $Z_{N}$ simple currents. In this sense, there exists some difference between the t'Hooft anomaly and anomaly of simple current.

So far, these two approaches, the RG domain wall interpretation of topological ordered systems and the modular property of the orbifolded theory seem different. In this work, we give a unified way to understand these phenomena by constructing the anomaly free composite $Z_{N}$ objects which have first appeared in the analysis of discrete torsion as the "integer spin simple current" by Schellekens and Gato-Rivera\cite{Gato-Rivera:1991bqv,Schellekens:1990ys}\footnote{For the review of this aspect of CFT with emphasis on the simple current, see \cite{Fuchs:1997af},for example.}. For this purpose, we first start from the $Z_{N}$ fractional quantum Hall states, because the mathematical procedure which corresponds to "gauging finite group symmetry" appears naturally\cite{Bhardwaj:2017xup,Tachikawa:2017gyf}.  This leads to the emergence of the modular $S$ invariant which has been expected to survive as protected edge modes\cite{Fukusumi_2022_f, Fukusumi_2022,Ino:1998by,Milovanovic:1996nj,Cappelli:1996np,2013PhRvX...3b1009L}. By considering the imaginary gapping operation (or gauging) of multicomponent bosonic parts, one can expect to obtain a class of general topological order\cite{Cheng:2022nji,Cheng:2022nds}. Similar proposals can be seen several condensed matter works \footnote{See for example \cite{Kong:2022cpy,Chatterjee:2022kxb} and reference therein.}, but by interpreting bulk edge correspondence as $CFT_{D}/BQFT_{D+1}$ correspondence, and its bulk RG flow as $CFT_{D}/BTQFT_{D+1}$ correspondence, one can expect a possibility of generalization of our analysis to both gapless and gapped models in general space-time dimensions $D$ or $D+1$. By gapping modular $S$ noninvariant parts, one may obtain modular $S$ invariant parts as candidates for protected edge modes. As we will show, this distinction between modular $S$ invariant part (or low energy part) and noninvariant part (or high energy part) for anomaly free models naturally leads to a dual model description of $Z_{N}$ models, which is analogous to fractional supersymmetry\cite{Ahn:1990gn,Durand:1992zk,Mohammedi:1994rm,Isaev:2001ni,RauschdeTraubenberg:1999tz,O_Brien_2020,Mong:2014ova,Perez:1996qc}. Moreover, in our discussion, one can understand the appearance of nonlocal CFT as a consequence of gauging anomalous symmetries.

The rest of the manuscript is organized as follows. In section \ref{cs_parafermion}, we revisit the integer spin simple currents and introduce them as composite particles, "composite parafermion", which can serve as a building block of the cylinder partition function of $Z_{N}$ FQHE. The anomaly freeness of these particles and its implication to the modular property are discussed. In section \ref{Shuffle}, we introduce the extension of our analysis for nonunitary theories and introduce a new mechanism, Galois shuffle cancelation. We introduce the case for $Z_{2}$, but applying the technique of the previous section and this section, one can construct the candidates of the protected edge modes of general nonunitary theories. In section \ref{duality_confinement}, we propose a new duality in the composite parafermion model, which can be interpreted as a kind of fractional supersymmetry. Its analogy with quark confinement is shown. Moreover, we also discuss a possible interpretation of the controversy around $\nu=5/2$ states\cite{Wang2017TopologicalOF,Mross_2018} with respect to composite Majorana fermionization and quantum anomaly. In section \ref{anomalous_pf}, we revisit the anomalous parafermionization in \cite{Yao:2020dqx} with emphasis on anomaly and chirality. The appearance of chirality or mirror symmetry breaking under the gauging anomalous symmetry is discussed, and its implications in a 2+1 dimensional system is shown. In section \ref{Haldane_cs_pf}, we give a concise understanding of $SU(N)$ Haldane conjecture based on composite parafermionization. In section \ref{conclusion}, we give a concluding remark, with emphasis on the analogy between FQHE and 2d quantum gravity coupled to matter. In the appendix, we show some detailed calculations in the main texts and also introduce the 1+1 dimensional analog of models with composite parafermion and confinement in the main text.

\section{Anomaly free $Z_{N}$ simple currents}
\label{cs_parafermion}
In this section, we introduce the anomaly free $Z_{N}$ simple current. For $N$ odd, this has already been proposed as integer spin simple current and is widely known because of its relation to discrete torsion\cite{Gato-Rivera:1991bqv,Schellekens:1990ys}. For $N$ even, as we will see, there exist two types of objects and they seem less well-known in the condensed matter theory \footnote{After submitting the manuscript, a work which revisits this anomaly free obeject appeared \cite{Kikuchi:2022ipr}.}.

First, let us assume the existence of $Z_{N}$ simple current generated by $J$ with minimal conformal weight $h_{J}$. The simple currents are represented as $J_{I}=J^{I}$ for $I=1,2, ...N-1$ with conformal dimension $h_{J_{I}}$ and they act like a group to the other fields in the theory. Then, the anomaly free object, which we call "composite parafermionic theory" satisfies the following conditions,
\begin{itemize}
\item{For $N$ odd, $h_{J_{I}}=\text{integer}$ for all $I$}
\item{For $N$ even, $h_{J_{I}}=\text{half-integer}$ for all $I$ odd and  $h_{J_{I}}=\text{integer}$ for all $I$ even, or $h_{J_{I}}=\text{integer}$ for all $I$ }
\end{itemize}

It should be noted that with these conditions, the simple currents satisfy,
\begin{equation}
Q_{J_{I}}(J_{I'})=0 \ (\text{mod.}1), \text{for all} \ I, I'
\end{equation}
where $Q_{J}(\alpha)=h_{\alpha}+h_{J}-h_{J\times \alpha}$ is the simple charge of an operator $\alpha$ (with comformal dimension $h_{\alpha}$) under the simple current $J$ (with confromal dimension $h_{J}$). To avoid complications, we do not specify the $(\text{mod}. 1)$ structure of the simple charges in the following discussion.
This is the generalized version of anomaly free condition in the LSM theorem\cite{Yao:2018kel}. In other words, the simple currents are untwisted from each other and they can condensate each other without anomaly obstructions in the anomaly free case. This phenomenon can be thought of as a $Z_{N}$ generalized version of BCS cooper pair creation known as anyon condensation\cite{Bardeen:1957mv,Kong2013AnyonCA,Neupert_2016,Eliens:2013epa,Burnell_2018}. To emphasize these objects satisfy the $Z_{N}$ fusion rule, we call them "composite parafermions" whereas they do not satisfy the parafermi statistics. In the later section, we also give the picture of these composite parafermions as composite particles of anomalous parafermions, known as the Fateev-Zamolodchikov parafermions\cite{Fateev:1985mm}. 

In the following discussion, we mainly consider characters of CFTs parametrized by the modular parameter $\tau$ and its complex conjugate $\overline{\tau}$. We mainly consider the following two modular transformations,
\begin{align}
&T:\tau \rightarrow\tau+1, \\
&S:\tau \rightarrow-\frac{1}{\tau},
\end{align}
and study the invariance and noninvariance of the partition functions under these transformations. As has been discussed in \cite{Fukusumi_2022_f, Fukusumi_2022,Ino:1998by,Milovanovic:1996nj,Cappelli:1996np}, one can expect the modular $S$ invariant parts of the total partition function can appear under bulk gapping operation with bulk-edge correspondence in cylinder geometry for 2+1 dimensional systems.

Here we note the partition function of the $Z_{N}$ composite parafermionic model constructed from bosonic partition function,  $Z_{M}=\sum_{i,p}\chi_{i,p}\overline{\chi}_{i,-p}+\sum_{a}|\chi_{a}|^{2}$where $i$ labels the $Z_{N}$ noninvariant sectors corresponding to the character $\chi_{i,p}$ and $\{p\}^{N-1}_{p=0}$ labels its parafermionic parity which takes an integer (mod.$N$) and $a$ labels the $Z_{N}$ invariant sectors corresponding to the character $\chi_{a}$ . The composite parafermionic partition function can be characterized by the simple charge of each sector, $\{ Q_{I}\}_{I=1}^{N-1}$, and is summarized as in the following form \footnote{The possible value of $\{ Q_{I}\}_{I=1}^{N-1}$ is restricted by the charge conjugate condition and detail of the models in general, but we do not specify the possible value to avoid complications of the discussion},

\begin{equation}
Z_{\{Q_{I}\}}=\sum_{i \in\{Q_{J_{I}}(i)=Q_{I}\}}|\sum_{p}\chi_{i,p}|^{2}+N\sum_{a\in\{Q_{J_{I}}(a)=Q_{I}\}}|\chi_{a}|^{2},
\label{CP_partition_function}
\end{equation}
where we have used the straightforward generalization of the discussion in \cite{Hsieh:2020uwb}. Because of anomaly freeness, $Q_{J_{I}}(i,p)$ does not depend on the parafermionic parity. Hence we have dropped the index of $p$ for the simple charges. Each charge sector may correspond to the twisted boundary conditions of the composite parafermionic chain, as we will show in the appendix. In unitary CFTs, one can obtain the sector with $Q_{I}=0$ for all $I$ as modular $S$ invariant and this plays a central role in the following construction of the $Z_{N}$ FQHEs. 

\subsection{$Z_{N}$ fractional quantum Hall effect}
\label{FQHE}

For a simple application, we study the fractional quantum Hall effect constructed from the $Z_{N}$ composite parafermion. We combine together the approaches in \cite{Schoutens:2015uia,Milovanovic:1996nj,Ino:1998by,Fukusumi_2022,Fukusumi_2022_f}. First, let us introduce the partition function of multicomponent bosonic fields $\varphi_{I}$ indexed by integer $q_{I}$, $I=1, 2, ...,N-1$ as,

\begin{align}
Z_{mb}&=\prod_{I=1}^{N-1}\sum_{r_{I}=0}^{q_{I}-1}\sum_{m_{I},\overline{m}_{I}}\frac{1}{|\eta(x)|^{2}} x^{\frac{(m_{I}q_{I}+r_{I})^{2}}{2q_{I}}}\overline{x}^{\frac{(\overline{m}_{I}q_{I}+r_{I})^{2}}{2q_{I}}}, 
\label{mb}
\\
x&=e^{2 \pi i\tau},\overline{x}=e^{2\pi i \overline{\tau}},
\end{align}
where  $\eta(x)=x^{-1/24}\prod_{n=1}^{\infty}(1-x^{n})$ is the eta function and $\tau$ and $\overline{\tau}$ are modular parameter corresponding to each edge, and $r_{I}$ corresponds to the number of quasihole. It should be noted that the $m_{I}$ and $\overline{m}_{I}$ correspond to the number of particles at each edge and take all integer values. In the following discussion, we take the summation indices $ r_{I}$, $m_{I}$ and $\overline{m}_{I}$ the same as in Eq.\eqref{mb} without a specific caution. 

The partition function \eqref{mb} corresponds to the direct product of the partition functions of Laughlin states with the filling factor $\{ 1/q_{I}\}^{N-1}_{I=1}$ and satisfies the modular $S$ and $T$ invariance. One can easily generalize the situation with other multicomponent bosonic models, such as the model with a general $K$ matrix.

For the later discussion, we introduce the following character,
\begin{equation}
\theta^{m_{I}}_{\frac{r}{q_{I}}}=\frac{1}{\eta(x)} x^{\frac{(m_{I}q_{I}+r_{I})^{2}}{2q_{I}}},
\end{equation}

Here we introduce the coupling of the multicomponent bosonic model and the CFT with composite parafermion. The general electronic objects generated by $Z_{N}$ simple currents are $\{J_{I}e^{i\sqrt{q_{I}}\varphi_{I}}\}^{N-1}_{I=1}$. Again, the simple charges of the operator indexed by $\alpha$ are,
 \begin{equation}
Q_{J_{I}}(\alpha)=h_{J_{I}}+h_{\alpha}-h_{J_{I}\alpha}.
\end{equation}
Corresponding to these twists, one has to redefine the chiral character $r_{I}\rightarrow r_{I}+Q_{J_{I}}(\alpha)$ for each primary field $\phi_{\alpha}$. This modification is analogous to the gravitational dressing or Knizhnik-Polyakov-Zamolodchikov (KPZ) relation in 2d quantum gravity coupled to matter\cite{David:1988hj,Knizhnik:1988ak,Distler:1988jt}, and FQHE and 2d quantum gravity coupled to matter share the basic theoretical structure, the single-valuedness of the correlation function. In other words, FQHE is a gauge theory analogous to string theory, and this similarity can be interpreted as a generalized (or relaxed) version of gauge/gravity duality. For the later discussion, we denote the character $\phi_{\alpha}$ without such twist, i.e. $Q_{J_{I}}(\alpha)=0$ for all $I$ as "untwisted". Besides these simple charges and the twists, one has to introduce $Z_{N}$ parity coming from the coupling of the models. Hence the resulting partition function becomes complicated when one only assumes the bulk-edge correspondence. However, because of the anomaly freeness, one can represent the resulting partition function with modular $T$ or $T^{2}$ invariance as,
\begin{equation}
Z=Z^{\text{untw}}+Z^{\text{tw}},
\end{equation}
where $Z^{\text{untw}}$ and $Z^{\text{tw}}$ correspond to the untwisted and twisted part of the theory classified by the simple charges $Q_{J_{I}}$. The explicit form of the above partition function is shown in appendix.

The detailed characters are summarized as,
\begin{align}
\Xi^{p' }_{i,p,\{r_{I}\}}&=\chi_{i,p}\prod_{I}\sum_{m_{I}; \sum_{I}m_{I}+p=p' (\text{mod}. N)}\theta^{m_{I}}_{\left(r_{I}+Q_{J_{I}}\left(\phi_{i,p}\right)\right)/q_{I}} , \\
\Xi_{a, \{r_{I}\}} &=\chi_{a}\prod_{I}\sum_{m_{I}}\theta^{m_{I}}_{\left(r_{I}+Q_{J_{I}}\left(\phi_{i,p}\right)\right)/q_{I}} ,
\end{align}
where ${i, p}$ and $a$ label $Z_{N}$ noninvariant and $Z_{N}$ invariant states respectively, and $p$ is the parafermionic parity coming from  parafermionic representation as we have introduced in the previous section. For the $Z_{N}$ invariant primary fields $\phi_{a}$ in the bosonic represntation, one can consider parafermionization as $\phi_{a}=\left(\sum_{p=0}^{N-1}\phi_{a,p}\right)/\sqrt{N}$. As in the $Z_{2}$ case, there exists gauge choices $\{ \nu_{p}\}_{p=0}^{N-1}$ as $\phi_{a}=\left(\sum_{p=0}^{N-1}\omega^{\nu_{p}}\phi_{a,p}\right)/\sqrt{N}$ where $\omega$ is $N$th root of unity, $\omega=\text{exp}\left( 2\pi i/N\right)$, but we have taken the simplest gauge that gives the nonnegative integer fusion rule. In analogy with $Z_{2}$ case, we call $\phi_{a,p}$ as $Z_{N}$ semion and we have assumed a counting rule which is a natural generalization as in \cite{Milovanovic:1996nj,Ino:1998by}. This may need further numerical tests as in \cite{Jolicoeur:2014isa}, because the modular property can change under this counting. The similar counting problem in $Z_{2}$ case has been discussed in \cite{Ino:2000wa,Lou:2020gfq}, for example.

By gapping fractional flux sectors, one can obtain the modular $S$ invariant partition function as,

\begin{equation}
\begin{split}
Z& \rightarrow Z^{\text{untw}}\\
 &=\left(\sum_{i, p\in \text{untwisted}}|\sum_{p=0}^{N-1}\chi_{i,p}|^{2}+N\sum_{a\in \text{untwisted}}|\chi_{a}|^{2}\right)Z_{mb}. 
\label{S_inv_FQHE}
\end{split}
\end{equation}
This is nothing but the product of Eq. \eqref{mb} and untwisted part of Eq.\eqref{CP_partition_function}, and satisfies the modular $S$ invariance.

Hence, for the $Z_{N}$ FQHE constructed from the composite parafermionic theories, there exists no obstruction to obtain bulk gappability from their modular $S$ property. In other words, we have obtained the candidates of the protected edge modes in the cylinder geometry. Next, we show the corresponding BCFT structure of the theories which has a close relationship with the bulk topological properties and the topological entanglement\cite{Fukusumi_2022_f,Qi_2012,Das:2015oha}. For simplicity, we drop off the indices coming from multibosonic parts in the rest of this section.

In this composite parafermionic representation, one can easily extend the analysis of fermionic BCFT in \cite{Fukusumi:2021zme,Weizmann} to anomaly free parafermionic BCFT. The results are,
\begin{align}
|i\rangle_{\text{PF}}&=\sum_{p=0}^{N-1} |i,p\rangle_{\text{Cardy}}, \\
|a,p\rangle_{\text{PF}}&=|a\rangle_{\text{Cardy}}\otimes |p\rangle, \\
\end{align}
where $i,p$ labels $Z_{N}$ noninvariant states and $a$ labels $Z_{N}$ invariant states in the bosonic fusion rule. We have introduced $|p\rangle$ as in the fermionic case to label $Z_{N}$ parafermionic parity\cite{Fukusumi:2021zme}.
Hence by mapping from the bulk degeneracy problem to the tunnel problem\cite{Wen:1990se}, and by applying Cardy's conjecture \cite{Cardy:2017ufe} to this situation, one can obtain the bulk topological degeneracies by considering the degeneracies of the untwisted particles, $\{\phi_{i,p}\}_{p=0}^{N-1}$ or $Z_{N}$ semion $\{\phi_{a,p}\}_{p=0}^{N-1}$\footnote{However, the way choosing the appropriate sector or the type of anyon may depend on the detail of the interactions corresponding to the detail of models. Some ambiguity of the bulk topological degeneracies may correspond to this\cite{Watanabe:2022pgk}. It may also be related to subtle relations between modular invariants and BCFTs \cite{Gannon:2001ki}.}. The $Z_{N}$ semions only appear as untwisted states when composite parafermion becomes integer spin simple current.

As one can see in Eq. \eqref{S_inv_FQHE}, the partition function naturally contains the subalgebra structure of fusion rule as the author has observed in the fermionic case\cite{Fukusumi_2022_f}, called Lagrangian subalgebra as,
\begin{equation}
\phi_{\alpha}^{\text{untw}}\times \phi_{\alpha'}^{\text{untw}} =\sum_{\beta} N_{\alpha,\alpha'}^{\beta} \phi_{\beta}^{\text{untw}},
\end{equation}
where $\alpha$, $\alpha'$, and $\beta$ are labeling untwisted fields (or extracting untwisted fields labeled by $\{ i,p\}_{p=0}^{N-1}$ and $\{ a,p\}_{p=0}^{N-1}$), and $N_{\alpha\alpha'}^{\beta}$ is the fusion matrix of the theory.
Moreover, the partition function Eq. \eqref{S_inv_FQHE} naturally represents the type of particles $\{ i,p \}^{N-1}_{p=0}$ or $\{ a,p \}^{N-1}_{p=0}$. By considering BCFTs, one can extract topological degeneracies corresponding to each type $i$ and $a$. 

Our construction can be applied to all anomaly free models constructed from the bulk-edge correspondence in principle. A typical example is $SU(N)_{K}$ FQHE constructed by Gepner parafermion\cite{Fuji_2017}. Hence we only introduce several examples and their implication in other aspects of physics in the next subsections.

\subsection{Coupling three states Potts three times }

One of the simplest models which  are in scope of the composite parafermionic $Z_{N}$ FQHE is the $Z_{3}$ model constructed by coupling three state Potts models three times and considering simple current construction as in\cite{Schoutens:2015uia}\footnote{Recently, such products of cyclic groups has captured some attentions of condensed matter physicists\cite{Topchyan:2022frk,Dupont2021FromTT}.}. Each Potts model has six primary fields, noted as $\{ I, \psi_{1},\psi_{2},\epsilon, \sigma_{1}, \sigma_{2} \}$, and $\psi_{1}$ and $\psi_{2}$ are the $Z_{3}$ simple currents. Where we have used the notation that parafermionic parity is identified in the lower index of the fields. The nontrivial fusion rule of each Potts CFT which is outside of $Z_{3}$ symmetry applications is,
\begin{align}
\epsilon \times \epsilon=I+ \epsilon .
\end{align}
This is nothing but the fusion rule of Fibonacci anyon.

In this coupled model, there exist $6^{3}=216$ primary fields. $J_{1}=\psi_{1}^{(1)}\psi_{1}^{(2)}\psi_{1}^{(3)}$ is a typical choice of the composite parafermion where the upper index corresponds to each copy of three state Potts CFT. The partition function by bulk edge correspondence contains a lot of fields, but if one considers the untwisted part which corresponds to protected edge modes, one can easily write down the cylinder partition function of the $Z_{3}$ FQHE model as, 
\begin{equation}
Z^{\text{untw}}_{(Z_{3})^{3}}=\sum_{p_{\alpha^{(1)}}=0, p_{\alpha^{(2)}}+p_{\alpha^{(3)}}=0} |\sum_{i=0}^{2}\chi_{J^{i}\alpha^{(1)}\alpha^{(2)}\alpha^{(3)}}|^{2}Z_{mb}
\end{equation}
where $\alpha^{(i)}$ takes $\{ I, \psi_{1},\psi_{2},\epsilon, \sigma_{1}, \sigma_{2} \}$ resepctively under the condition $\{ p_{\alpha^{(1)}}=0, p_{\alpha^{(2)}}+p_{\alpha^{(3)}}=0\}$, where $p_{\alpha^{(i)}}$ represents the parity of the field $a^{(i)}$. It contains $(2\times 6\times 2)\times 3=72$ primary fields coming from coupled Potts part of CFT.

In this model, there exists the other anomaly free $Z_{3}$ simple currents with total parafermionic parity $1$, $J_{1}'=\psi_{1}^{(1)}\psi_{1}^{(2)}\psi_{2}^{(3)}$ and their permutations by the upper indices. The partition function for this case is,
\begin{equation}
Z^{\text{untw}}_{(Z_{3})^{3}}=\sum_{p_{\alpha^{(1)}}=0, p_{\alpha^{(2)}}-p_{\alpha^{(3)}}=0} |\sum_{i=0}^{2}\chi_{J^{i}\alpha^{(1)}\alpha^{(2)}\alpha^{(3)}}|^{2}Z_{mb}
\end{equation}

Similarly, one can consider anomaly free $Z_{N}$ coupled Potts model and the corresponding FQHE by coupling Potts model $KN$ times for odd $N$ and $KN/2$ times for even $N$ where $K$ is an arbitrary positive integer. As one can easily see, this type of coupled model has a gigantic number of primary fields, approximately $N^{ANK}\prod_{I} q_{I}$ where $A$ is a constant depending on the model.

\subsection{$SU(N)$ spinon-holon state from $SU(N)_{K}$ Wess-Zumino-Witten model}

In $SU(N)_{K}$ Wess-Zumino-Witten models, the $Z_{N}$ simple currents $J_{I}$ have the following conformal dimensions,
\begin{equation}
h_{J_{I}}=\frac{KI(N-I)}{2N}, I=1,...,N-1
\end{equation}
Hence $SU(N)_{K}$ WZW models have the composite parafermionic representation when $K=0 \ (\text{mod}.N)$, and one can consider gapped FQHE by coupling $N-1$ multicomponent bosons. This can be thought of as a generalization of spinon-holon states proposed in \cite{Ino:1998by}. It is interesting to note that even when $K$ does not satisfy the composite parafermion condition, it may be possible to construct modular $S$ invariant partition functions by extending the method in \cite{Ino:1998by}. However, these partition functions break modular $T$ and $T^{2}$ invariance and they may not have the corresponding lattice model in 1+1 dimensional quantum systems with local interaction. Related discussion under the thin torus limit of anomalous (gapless) FQHE model can be seen in \cite{Papi__2014}.

\section{Anomaly and Galois shuffle cancelation in coupled nonunitary models}
\label{Shuffle}

In this section, we discuss a possibility to generalize our discussion to nonunitary CFTs. For simplicity, we concentrate on a $Z_{2}$ symmetric case.
In \cite{Davenport_2013}, the spin singlet Gaffnian state which is a natural generalization of the Gaffnian FQH state has been proposed. Their model can be described by the product of nonunitary minimal models, $M(3,r+2)\times M(r,2r+1)$ and chiral Lattinger liquid. Interestingly, the product theory is closely related to the RG domain wall in \cite{Gaiotto:2012np} when $r=3$. In this section, we revisit the structure of this type of modelswith the view of anomaly defined by the simple charge coming from the operators and monodromy charge coming from modular $S$ property of quantum states. Consequently, we introduce the notion of Galois shuffle cancelation (shuffle cancelation in short) which represents the cancelation of anomalous monodromy charge.

In nonunitary $Z_{2}$ anomalous rational minimal model $M(p,q)$, there exist two remarkable properties:
\begin{itemize}
\item{The conformal dimension $h_{J}$ of $Z_{2}$ simple current $J$ takes $h_{J}=(2k+1)/4$ and becomes anomalous where $k$ is an integer.}
\item{The simple charge $Q_{J}(\alpha)$ of the arbitrary operator $\phi_{\alpha}$  does not match with the monodromy charge $Q^{\text{mon}}_{J}(\phi_{\alpha})=Q_{J}(\phi_{\alpha})-Q_{J}(\sigma)$ of the same operator where the $\sigma$ is the operator with lowest conformal dimension and $Q_{J}(\sigma)=\text{half-integer}$.}
\end{itemize}
The former property defines the anomalous $Z_{2}$ symmetry and the latter defines the nonunitarity of the model. The later property of the shuffle between the simple charge and the monodromy charge has been called Galois shuffle\cite{Gannon:2003de}. We name the operator $\sigma$ as shuffle generator, for convenience. In considering FQHE, the former becomes relevant to consider modular $T$ property or locality of the model and the latter becomes relevant to consider modular $S$ property or stability of the edge modes\cite{Ino:1998by}. Here we mostly discuss the later property. In FQHE, whereas the wavefunction is constructed by considering correlation functions and corresponding simple charge, the modular property coming from gauging $Z_{2}$ symmetry structure is defined by the monodromy charge. This difference between monodromy charge (or the symmetry charge of the states) and simple charge (or the symmetry charge of operators) has been studied extensively in \cite{Gannon:2003de} and stressed at least in two different fields, mathematical physics as in \cite{Harvey:2019qzs} and earlier analysis of Haldane-Rezayi model\cite{Milovanovic:1996nj,Ino:1998by}. 

Here, let us introduce the coupling of the two anomalous nonunitary conformal field theories $M_{1}$, $M_{2}$ with $Z_{2}$ simple current $J^{(1)}$ and $J^{(2)}$, and with Galois shuffle by the lowest conformal dimension fields $\sigma^{(1)}$ and $\sigma^{(2)}$  respectively. In this setting, the coupled theory has a nonanomalous $Z_{2}$ simple current $J^{(1)}J^{(2)}$. Moreover, the shuffle property of the product theory is canceled as we show in the following.

One can calculate the monodromy charge of the operator $\phi_{\alpha^{(1)}}\phi_{\beta^{(2)}}$ in the product theory as,
\begin{equation}
\begin{split}
& Q_{J_{1}J_{2}}^{\text{mon}}(\phi_{\alpha^{(1)}} \phi_{\beta^{(2)}}) \\
&Q_{J^{(1)}J^{(2)}}(\phi_{\alpha^{(1)}}\phi_{\beta^{(2)}})-Q_{J_{1}J_{2}}(\sigma^{(1)}\sigma^{(2)}) \\
&=Q_{J^{(1)}}(\phi_{\alpha^{(1)}})-Q_{J^{(1)}}(\sigma^{(1)})+Q_{J_{2}}(\phi_{\beta^{(2)}})-Q_{J^{(2)}}(\sigma^{(2)}) \\
&=Q_{J^{(1)}}(\phi_{\alpha^{(1)}})+Q_{J^{(2)}}(\phi_{\beta^{(2)}}) \\
&=Q_{J_{1}J_{2}}(\phi_{\alpha^{(1)}} \phi_{\beta^{(2)}}),
\end{split}
\end{equation}
where we have used the shuffle property $Q_{J^{(1)}}(\sigma^{(1)})+Q_{J^{(2)}}(\sigma^{(2)})=0$. Hence the product theory $M_{1}\times M_{2}$ shows cancelation of anomaly and Galois shuffle.

Because of this cancelation of anomaly and shuffle, one can construct modular $S$ invariant fermionic theory and the corresponding fermionic FQHE as in usual nonanomalous unitary CFT by taking $J^{(1)}J^{(2)}$ as $Z_{2}$ simple current\cite{Cappelli:2010jv,Fukusumi_2022_f}. Hence, at least from this bulk-edge correspondence, the spin singlet Gaffnian state can be gapped out under preserving bulk-edge correspondence, whereas its gaplessness has been proposed in \cite{Davenport_2013}. Further study and construction of bulk perturbation in various geometries in the lattice models are necessary to show their bulk gap or gaplessness. One can easily apply the same discussion to more general $Z_{N}$ models which can include both unitary and nonunitary CFT.

\section{Duality and confinement of composite parafermion models coming from anomalous parafermions}
\label{duality_confinement}
In this section, we propose a (hopefully new) $Z_{N}$ duality which is a $Z_{N}$ generalization of supersymmetry known as fractional supersymmetry \cite{Mohammedi:1994rm} or composite parafermionic T duality as in the $Z_{2}$ models\cite{Fukusumi_2022_f}.
As a simple model, we study the anomaly cancelation of $SU(3)_{1}$ WZW model and three state Potts model.

In $SU(3)_{1}$ WZW model, there exist three pimary fields, $\{ I, j_{1}, j_{2}\}$ with conformal dimension $h_{I}=0, h_{j_{1}}=1/3, h_{j_{2}}=1/3$ where $j_{1}$ and $j_{2}$ are the $Z_{3}$ simple currents.

As in the previous subsections, one can construct the two sets of composite parafermions, $\{ I, \psi_{1}j_{1}, \psi_{2}j_{2} \}$ or  $\{ I, \psi_{2}j_{1}, \psi_{1}j_{2} \}$. Each representation has $6$ primary fields as untwisted sectors and one can expect that they appear as protected edge modes of the $Z_{3}$ FQHE constructed from these composite parafermions.

Here we comment on the duality of the model. For simplicity, let us consider the situation when the composite parafermion is $\{ I, \psi_{1}j_{1}, \psi_{2}j_{2} \}$. Then we can obtain the set of twisted and untwisted partition functions corresponding to the simple charge $Q_{1}=Q_{j_{1}\psi_{1}}=0, \ 1/3, \ 2/3$ as,
\begin{align}
Z_{Q_{1}=0}&= |\chi_{I}+\chi_{j_{1}\psi_{1}}+\chi_{j_{2}\psi_{2}}|^{2}+ |\chi_{\epsilon}+\chi_{j_{1}\sigma_{1}}+\chi_{j_{2}\sigma_{2}}|^{2}, \\
Z_{Q_{1}=1/3}&= |\chi_{j_{1}}+\chi_{j_{2}\psi_{1}}+\chi_{\psi_{2}}|^{2}+ |\chi_{j_{1}\epsilon}+\chi_{j_{2}\sigma_{1}}+\chi_{\sigma_{2}}|^{2},\\
Z_{Q_{1}=2/3}&= |\chi_{j_{2}}+\chi_{\psi_{1}}+\chi_{j_{1}\psi_{2}}|^{2}+ |\chi_{j_{2}\epsilon}+\chi_{\sigma_{1}}+\chi_{j_{1}\sigma_{2}}|^{2},\\
\end{align}
where we have used the charge conjugate condition and dropped off the label coming from $Q_{2}=Q_{j_{2}\psi_{2}}$.

As can be seen from these partition functions, one can obtain this set of partition functions starting from the untwisted partition function $Z_{Q_{1}=0}$ by shifting the parity or simple charge by the operator $j_{1}$. This duality generated by the anomalous parafermionic parity shift or simple charge shift can be thought of as a kind of $Z_{N}$ version of supersymmetry, called fractional supersymmetry\cite{Ahn:1990gn,Durand:1992zk,Mohammedi:1994rm,Isaev:2001ni,RauschdeTraubenberg:1999tz,O_Brien_2020,Mong:2014ova}. One can name this general duality as "composite parafermionic T duality". In the existing literature, the anomalous $Z_{N}$ particles with fractional spin and the associated currents have been considered as a characteristic property of fractional supersymmetric models. However, the composite parafermionic T duality is a consequence of anomaly freeness. Related anomaly free representation can be seen, for example in \cite{Perez:1996qc}. By applying the general construction of $Z_{3}$ FQHE in the cylinder geometry, one can observe that the only untwisted sector can survive as the edge modes and the other twisted sectors go to the bulk excitations. In this understanding, it should be stressed that the anomalous particle, $j_{i}$, $\psi_{i}$, and $\sigma_{i}$ do not appear without anomaly cancellation or "paring". This is analogous to quark confinment, because such anomalous particles (or "quarks") go to high energy states under the bulk gapping process. Hence by applying the bosonization to our model, the anomalous particle may appear, and the fractional supersymmetry in the literature can appear naturally. In this sense, it seems appropriate to interpret our composite parafermionic T duality as a more primitive version of fractional symmetry.

One can apply a similar analysis to more general $Z_{N}$ models by coupling anomalous $Z_{N}$ models. For the simplest model let us introduce the partition function of $\{ SU(N)_{1}\}^{N}$ model with $J_{I}=\prod_{k}j^{(k)}_{I}$ as composite parafermion where the upper index $(k)$ specifies each copy of $SU(N)_{1}$. In this model, each $SU(N)_{1}$ has only simple currents and identity fields, $I, j_{1}, ...,j_{N-1} $ as primary fields. The untwisted partition function is,
\begin{equation}
Z^{\text{untw}}=\sum_{\sum p_{\alpha^{(i)}}=0}\left|\sum_{I=0}^{N-1} \chi_{J_{I}\prod_{k=2}^{N}\alpha^{(k)}}\right|^{2}.
\end{equation}
where $p_{\alpha^{(k)}}$ is the parafermionic parity of the operator $\alpha^{(k)}$ and $\alpha^{(k)}$ takes $I, j_{1}, ...,j_{N-1} $ respectively.
The twisted part can be obtained by adding $j^{(1)}_{1}$ recursively for example. Hence one can observe the analogy of quark confinement even in this simple model.

\begin{figure}[htbp]
\begin{center}
\includegraphics[width=0.5\textwidth]{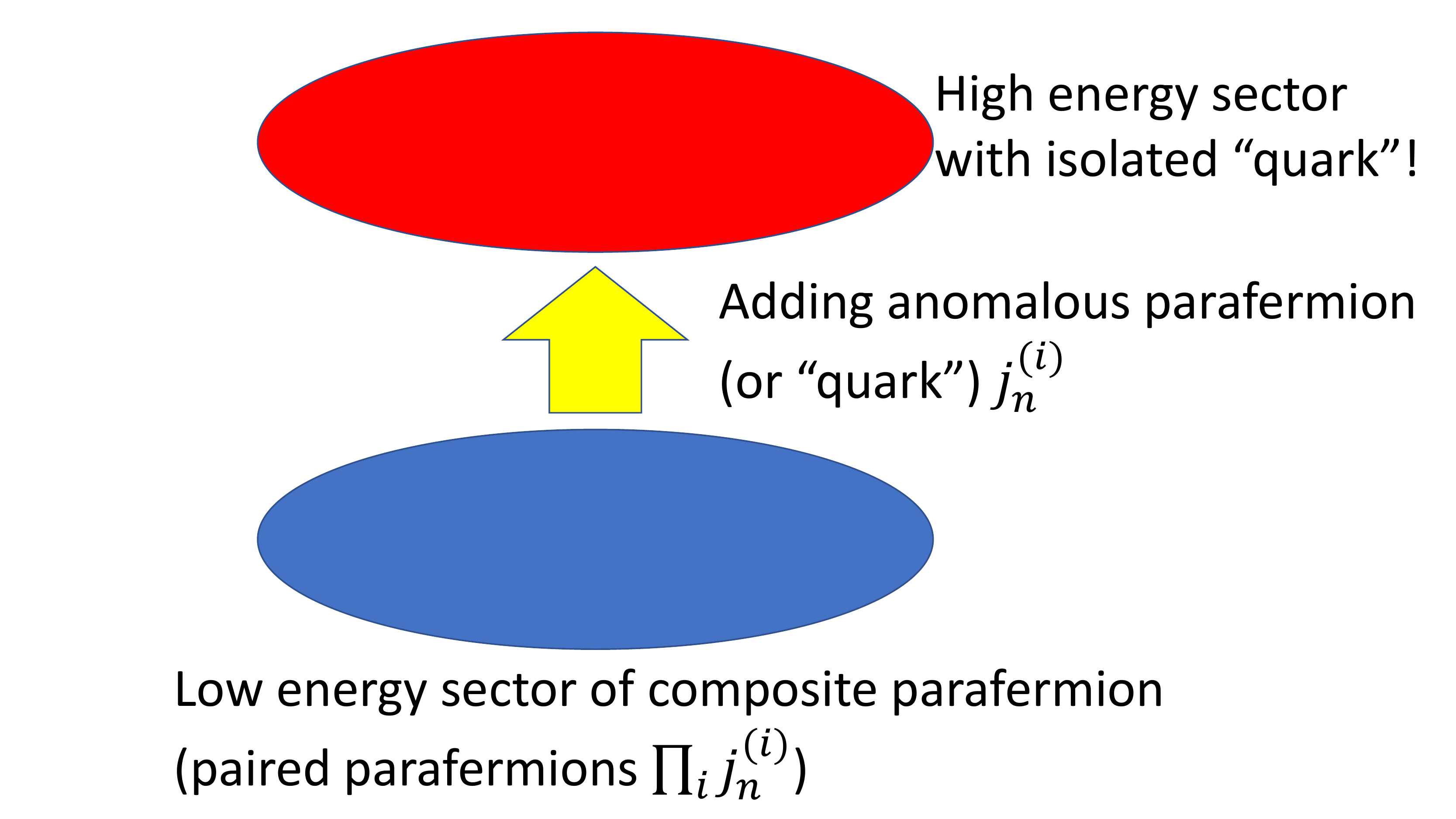}
\caption{Analogy between quark confinement and FQHE}
\end{center}
\end{figure}

It may be interesting to consider the relation between $Z_{N}$ duality in the general coset CFTs and such $Z_{N}$ composite parafermionic T duality of the coupled models. For the simples example, let us consider $SU(2)_{1}\times SU(2)_{1}=M(3,4)\times SU(2)_{2}$. The equivalence is ensured by coset construction of the Ising model. Related discussion can be seen in \cite{Thorngren:2021yso}, for example. The primary fields of the lefthand side are the tensor product of two $SU(2)_{1}$ whose primary fields are $I, \phi_{1}$, with conformal dimension $h_{I}=0, h_{\phi_{1}}=1/4$, and its fusion rule is just $Z_{2}\times Z_{2}$. The anomaly free $Z_{2}$ is generated by $I$ and $\phi^{(1)}_{1}\phi^{(2)}_{1}$ where the upper index is labeling the copy of $SU(2)_{1}$. The other two primary fields $\phi^{(1)}_{1}$, $\phi^{(2)}_{1}$ are anomalous. Hence these anomalous $Z_{2}$ field generates fermionic T duality. The NS and R sectors of fermionic partition functions are obtained as,
\begin{align}
Z_{NS}&=|\chi_{0}+\chi_{\phi^{(1)}_{1}\phi^{(2)}_{1}}|^{2}\\
Z_{R}&=|\chi_{\phi^{(1)}_{1}}+\chi_{\phi^{(2)}_{1}}|^{2}
\end{align}
 On the other hand, in the representation $M(3,4)\times SU(2)_{2}$, the fusion rule of $M(3,4)$ and $SU(2)_{2}$ are the same Tambara-Yamagami category. The $Z_{2}$ duality operator has conformal dimension $h_{\sigma}=1/16$ and $h_{\phi_{1}}=3/16$ and the product of these fields, $\sigma \phi_{1}$ has conformal dimension $h_{\sigma\phi_{1}}=1/4$. Hence one can observe the product of the two $Z_{2}$ Kramers-Wannier duality becomes anomalous $Z_{2}$ parity shift. This picture is consistent with the intuitive understanding of semionic statistics of duality operators. We expect this mechanism is ubiquitous and gives some understanding of RG and anomaly in the way analogous to "particle" or "quark" physics. In appendix, we show a general construction of such a quark confinement-like model in 1+1 dimension from bosonic CFT and the corresponding lattice model.

As another example, one can consider anomaly analysis of the double semion model, which is equivalent to the fermionized fusion rule of $M(3,4)$ model. In this model there exist four fields $\{ I, \psi, e, m\}$ with the conformal dimensions $h_{I}=0, h_{\psi}=1/2, h_{e}=h_{m}=1/16 $. Except for the identity operator,  all fields in this theory can be thought of as $Z_{2}$ simple current, but $e$ and $m$, known as semion, have anomalous conformal dimension $1/16$. Hence one can expect to obtain anomaly free theory coupling this semion model $16$ times\footnote{To obtain the theory with integer or half integer simple charge, $4$ semion might be enough. Hence there may exist $4$-fold periodicity, more generally. This analysis can be extended to other $Z_{2}$ and $Z_{N}$ models.}. This may give an intuitive understanding of Kitaev's $16$-fold periodicity\cite{Kitaev:2006lla}.

\subsection{Hierarchy from coupled Majorana fermion and its implication on bulk and boundary flows}

\begin{figure}[htbp]
\begin{center}
\includegraphics[width=0.5\textwidth]{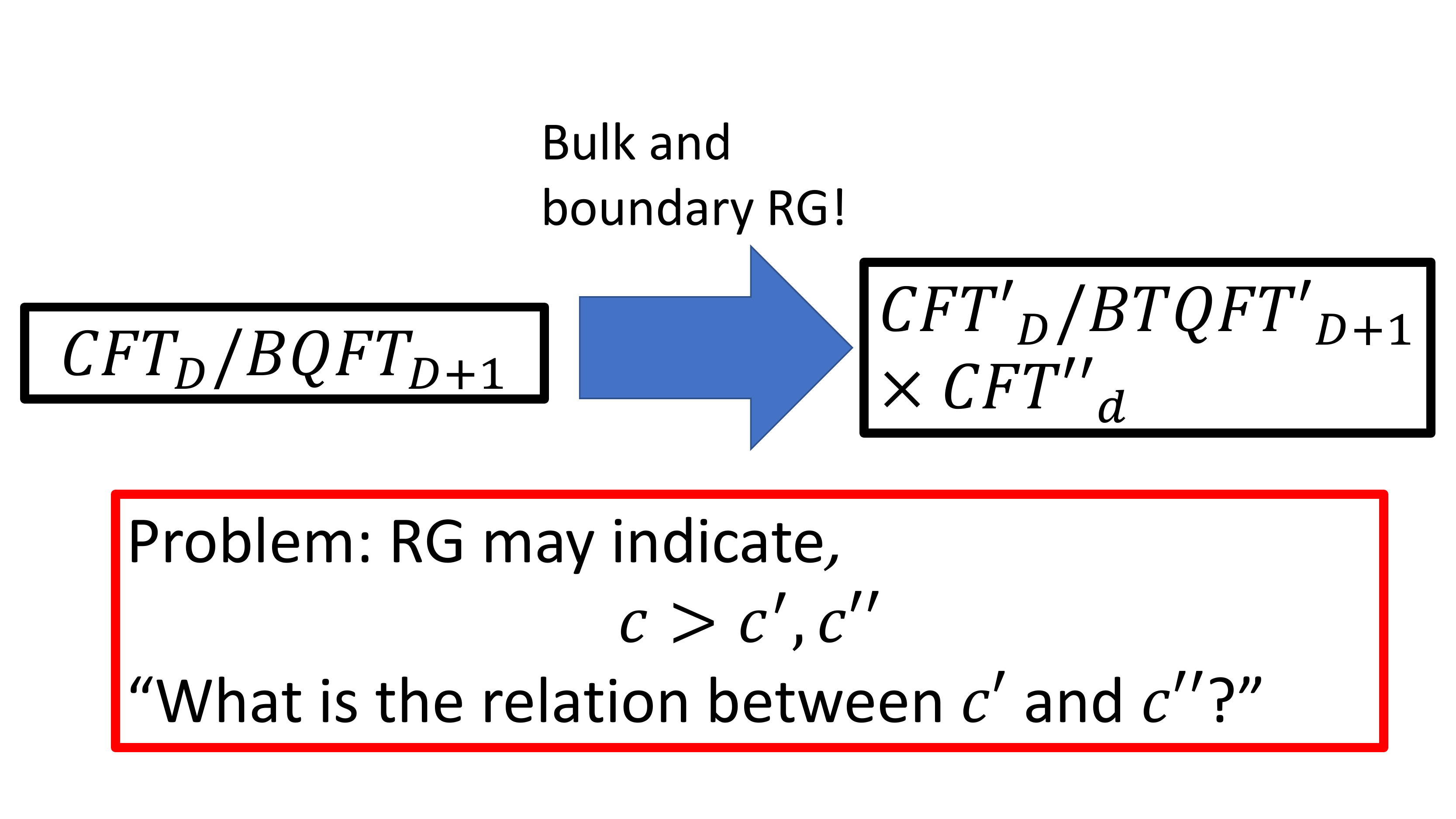}
\caption{Bulk and boundary RG interpretation of CFT/BQFT correspondence and its ambiguities. In a general topological order, one can attach different CFTs connected under RGs at the edge theoretically. This nonuniqueness of edge theory may result in difficulty in observing edge CFT in experimental settings.}
\label{bqft}
\end{center}
\end{figure}

Interestingly, as we have discussed, $M(3,4)$ and $SU(2)_{2}$ models have the same fusion rule. Moreover, it is well-known that there exists three components Majorana fermionic representation for $SU(2)_{2}$ WZW model. It coincides with a generalized anomaly cancelation mechanism which is analogous to the periodicity of spin-$S$ $SU(2)$ Heisenberg model in \cite{MOshikawa_1992}. As a generalization, one can consider five component fermionic theory. In this line, the corresponding central charges of the FQHE model constructed from $M(3,4)$, $SU(2)_{2}$, and $\{M(3,4)\}^{5}$ take $1/2+1$, $3/2+1$, $5/2 +1$ and this seems to correspond to the controversy about experimental and numerical observation of Moore-Read states in \cite{Mross_2018,Wang2017TopologicalOF}. It might be interesting to note that the number $5$ is the smallest number which becomes coprime with the ``conductor", $48$ of $M(3,4)$ in \cite{Harvey:2019qzs}. Hence following the approaches in \cite{Harvey:2019qzs}, $\{ M(3,4)\}^{5}$ may have the same fusion rule with $M(3,4)$ by considering its orbifolding\footnote{ Interestingly, the Hecke operation which is a central structure of their construction is similar to Renyfication which is closely related to the entanglement of models. With respect to RG, it may be interesting to consider their construction and cristalization of operators under Renyification \cite{St_phan_2009}.}. It should be noted that all these theories can be connected to each other without anomaly and spin statistics obstruction. In bulk and boundary RG, it may indicate a nontrivial relation between CFT which constructs the bulk wavefunctions (with central charge $c'$) and appears at the boundary with central charge $c''$ probably with $c''>c'$.  If one starts from $D+1$ dimensional bulk-edge correspondence or $CFT_{D}/BQFT_{D+1}$ correspondence with sufficiently large central charge $c$, or with sufficient degrees of freedom, one can safely explain the relation between these CFTs as the consequence of bulk and boundary RG flow, with $c>c'$, $c>c''$ (FIG. \ref{bqft}). However, it may be difficult to determine this $c$ started from a lattice model or experimental setting whereas it seems possible to construct such bulk and boundary RG flow purely from QFT. This difficulty is coming from the absence of a general criterion to determine the sufficiency of degrees of freedoms\footnote{As the author has stressed in \cite{Fukusumi:2020irh}, $BQFT_{D+1}$ can contain unstable boundary conditions which are outside of Cardy's condition}. Hence further studies of bulk and boundary RG flows are necessary to establish a general theory of protected edge modes.

\section{Anomalous parafermionization}
\label{anomalous_pf}
In this section, we introduce a field theoretic analog of parafermionization of $1+1$ dimensional quantum systems in the existence of $Z_{N}$ anomaly and its implication to $2+1$ dimensional systems constructed by the bulk-edge correspondence. We mainly follow the approach by Yao and Furusaki\cite{Yao:2020dqx} and express their results in a more unified way by using gauging. Studies of $Z_{2}$ anomalous symmetry can be seen in several related works \cite{Hao:2022kxo,Chatterjee:2022tyg,Chatterjee:2022kxb}.

First, let us introduce the paraspin variables at each site $l$ satisfying the following relations,
\begin{equation}
\sigma^{N}_{l}=\tau^{N}_{l}=1, \ \sigma_{l}^{\dagger}=\sigma_{l}^{-1}, \ \tau_{l}^{\dagger}=\tau_{l}^{-1},  \ \sigma_{l}\tau_{l}=\omega \tau_{l}\sigma_{l},  
\end{equation}
where $\omega=\text{exp}(2 \pi i /N)$
A typical example is the $Z_{N}$ Fateev-Zamolochikov spin chain and its parafermionization \cite{Albertini:1993hb} which can be described by $Z_{N}$ parafermion CFT\cite{Fateev:1985mm}. This type of parafermionic model has captured the attentions of condensed matter physicists as a new type of topological phase. These models have been expected to have a possibility to realize the topological quantum computation because of the existence of nonabelian anyons in the theories\cite{Santos_2017, Alicea:2015hja}. 

The spin chain Hamiltonian of $Z_{N}$ Fateev Zamolochikov model with chain length $L$ is,
\begin{equation}
H_{FZ}=-\sum_{l=1}^{L}\sum_{n=1}^{N-1}\frac{1}{\text{sin}\left(n\pi /N\right)}\left( \sigma^{-n}_{l}\sigma^{n}_{l+1}+\tau_{l}\right)
\end{equation}

Then, let us introduce the order and disorder fields in the $Z_{N}$ models following the existing works, \cite{Yao:2020dqx}. 
The disorder operator is
\begin{equation}
\mu_{l}=\prod_{j=1}^{l-1} \tau_{j},
\end{equation}
Then, one can define the parafermionic operator as,
\begin{align}
\gamma_{2l-1}&=\sigma_{l}\mu_{l}  \\
\gamma_{2l}&=\omega^{(N-1)/2}\sigma_{l}\mu_{l}
\end{align}

There exists the other choice of disorder operator \cite{Mong:2014ova} as,
\begin{equation}
\mu'_{l}=\prod_{j=L}^{l} \tau_{j},
\end{equation}
This gives the other parafermionization,
\begin{align}
\gamma_{2l}'&=\sigma_{l}\mu'_{l}  \\
\gamma_{2l-1}'&=\omega^{(N-1)/2}\sigma_{l}\mu'_{l}
\end{align}

These parafermionizations result in different chirality,
\begin{align}
\gamma_{l}\gamma_{l'}=\omega^{\text{sgn}(l-l')}\gamma_{l}\gamma_{l'}, \\
\gamma'_{l}\gamma'_{l'}=\omega^{-\text{sgn}(l-l')}\gamma'_{l}\gamma'_{l'},
\end{align}

\begin{figure}[htbp]
\begin{center}
\includegraphics[width=0.5\textwidth]{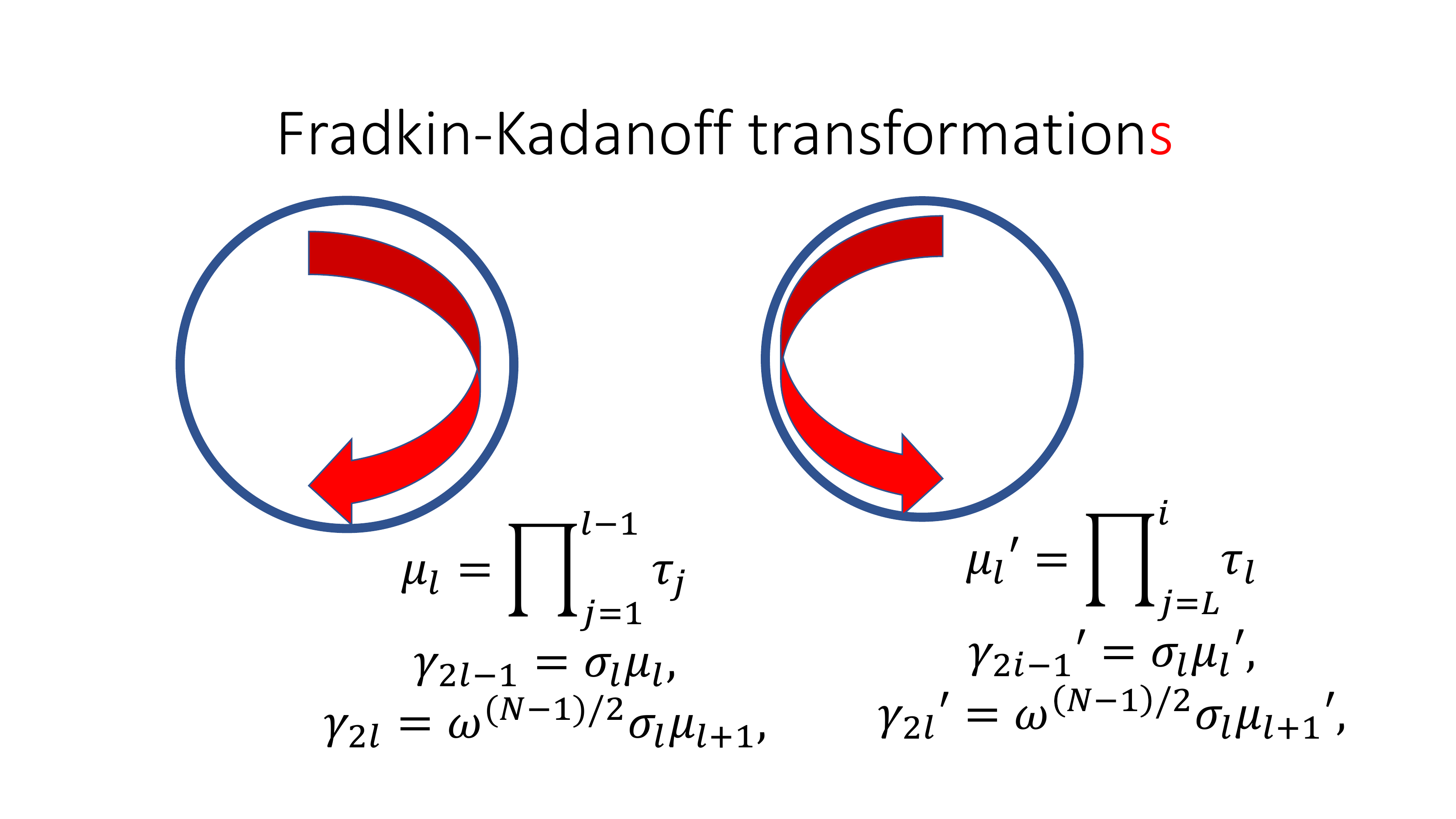}
\caption{Chirality from Fradkin-Kadanoff transformations}
\end{center}
\end{figure}

Because of the anomaly, the two parafermionizations do not give the same algebraic relations whereas the models have been constructed from the identical model. Because of this mismatch,  the chiralities of the theories appear. One can observe this more generally by considering the parafermionic partition functions in \cite{Yao:2020dqx}, and their non-invariance under the mirror transformation $\tau \leftrightarrow \overline{\tau}$. 

For the chiral parafermionic model, one can summarize the partition function characterized by the simple charge $\{ Q_{I}\}_{I=1}^{N-1}$ as,

\begin{equation}
\begin{split}
&Z_{\{Q_{I}\}} \\
&=\sum_{i, p\in\{Q_{J_{I}}(i,p)=Q_{I}\}}\left(\chi_{i,p}\right)\left(\sum_{p'}\overline{\chi}_{i,p'}\right),
\end{split}
\end{equation}
where we have assumed the bosonic modular invariant as $\sum_{i,p}\chi_{i,p}\overline{\chi}_{i,-p}$ and this does not have $Z_{N}$ invariant sector denoted as $\{a\}$ in the previous sections because of the quantum anomaly. By replacing $\tau$ and $\overline{\tau}$, one can obtain the antichiral parafermionic model\footnote{It might be interesting to note that summation of the chiral and antichiral partition functions satisfy the usual modular $S$ property as in the $Z_{2}$ case\cite{Fukusumi_2022}, and corresponds to bosonic twisted partition function generated by Verlinde line. This may be related to anomalous topological order, for example in \cite{Kong:2014qka}, but it is out of the scope of the present work.}.

In this sense, the space inversion symmetry or mirror symmetry of the parafermionic theory is broken by the anomaly. The action of mirror operation in rational CFT has been studied in \cite{Brunner:2003zm} for example, and our observation may clarify the relationship between chirality and anomaly. This nonuniqueness has appeared in \cite{Mong:2014ova} for example, but we cannot find the works discussing its relation to the nonuniqueness of partition function and its interpretation by quantum anomaly except for the author's other related work\cite{Fukusumi_2022}. This nonuniqueness may be trivial to some extent by considering the meaning of quantum anomaly in a textbook or its historical aspects\cite{Harvey:2005it}. However, we would emphasize its importance in considering the modular property and locality of the theory. 

To construct FQH systems from anomalous theory, it is necessary to introduce the flux part nontrivially to cancel $Z_{N}$ anomaly of the model. This coupling condition complicates the structure of the partition function in a quite nontrivial way. In this type of modular $S$ invariant, it is observed that the left and right edges belong to the different flux sectors\cite{Cappelli:1996np,Ino:1998by}. In this model, it is difficult to construct BCFT and the theory seems to be outside of the usual local quantum field theory by using the same label of fields. Interestingly, by introducing the nontrivial identification between flux sector to rational CFT, such as the mapping between $SU(N)_{1}\sim \{ U(1)\}^{N-1}$, it becomes possible to obtain a local QFT and the corresponding BCFT. The model $Z_{3}\times SU(3)_{1}$ is a typical example, by idnetifying the $SU(3)_{1}$ part as flux. However, whereas the untwisted part of $Z_{3}\times SU(3)_{1}$ corresponds to FQHE, their twisted part does not correspond to FQHE. Related observations can be seen in \cite{Papi__2014}, and recent research on anomalous symmetry can be seen in \cite{Hao:2022kxo,Chatterjee:2022tyg,Chatterjee:2022kxb}.

\section{$SU(n)$ Haldane conjecture and composite parafermionization}
\label{Haldane_cs_pf}
In this section, we revisit $SU(N)$ Haldane conjecture, with emphasis on quantum anomaly and composite parafermion condensation. Related discussion can be seen in \cite{Yao:2018kel} with emphasis on LSM anomaly and in \cite{Lecheminant:2015iga} with emphasis on integrability and coset representations. 

As has been proposed in \cite{Fuji_2017,Lecheminant:2015iga}, there exists mapping from $\{SU(N)_{1}\}^{K}$ to $SU(N)_{K}$. Remarkablely, the mutual simple charges from the simple currents $\{ \prod_{k=1}^{K}j_{n}^{(k)} \}^{N-1}_{n=1}$ in the $\{SU(N)_{1}\}^{K}$ and those of $\{ J_{I}\}_{I=1}^{N-1}$ in $SU(N)_{K}$ have the same structure. This can be interpreted as anomaly matching and this should survive under RG flow. Hence, by analyzing the simple theory $\{SU(N)_{1}\}^{K}$ constructed by simply taking products of $SU(N)_{1}$, one can expect some information of RG connectivity of CFT with a smaller central charge, and this corresponds to the $SU(N)$ Haldane conjecture (FIG.\ref{Haldane}). The commutativity of the diagram in FIG.\ref{Haldane} can be interpreted as Haldane conjecture. Under the RG flows of the red arrows, the simple currents anomaly which can be considered as the on-site symmetry charges, gradually become the LSM-like anomalies. Hence the RG connectivity of $\{ U(1)\}^{N-1}$ model with some anisotropy parameters as in XXZ chain, can be changed depending on simple charge anomaly of the product theory with a larger central charge. Typically, one may expect to observe the connectivity problem to $SU(N)_{1}$ WZW models\cite{Yao:2018kel}. It should be worth noting that this type of product theory has appeared in the analysis of discrete torsion theory\cite{Gato-Rivera:1991bqv}, and it has a close relation with integrable perturbations known as $\lambda$ deformation\cite{Georgiou:2017jfi,Georgiou:2018gpe}.

\begin{figure}[htbp]
\begin{center}
\includegraphics[width=0.5\textwidth]{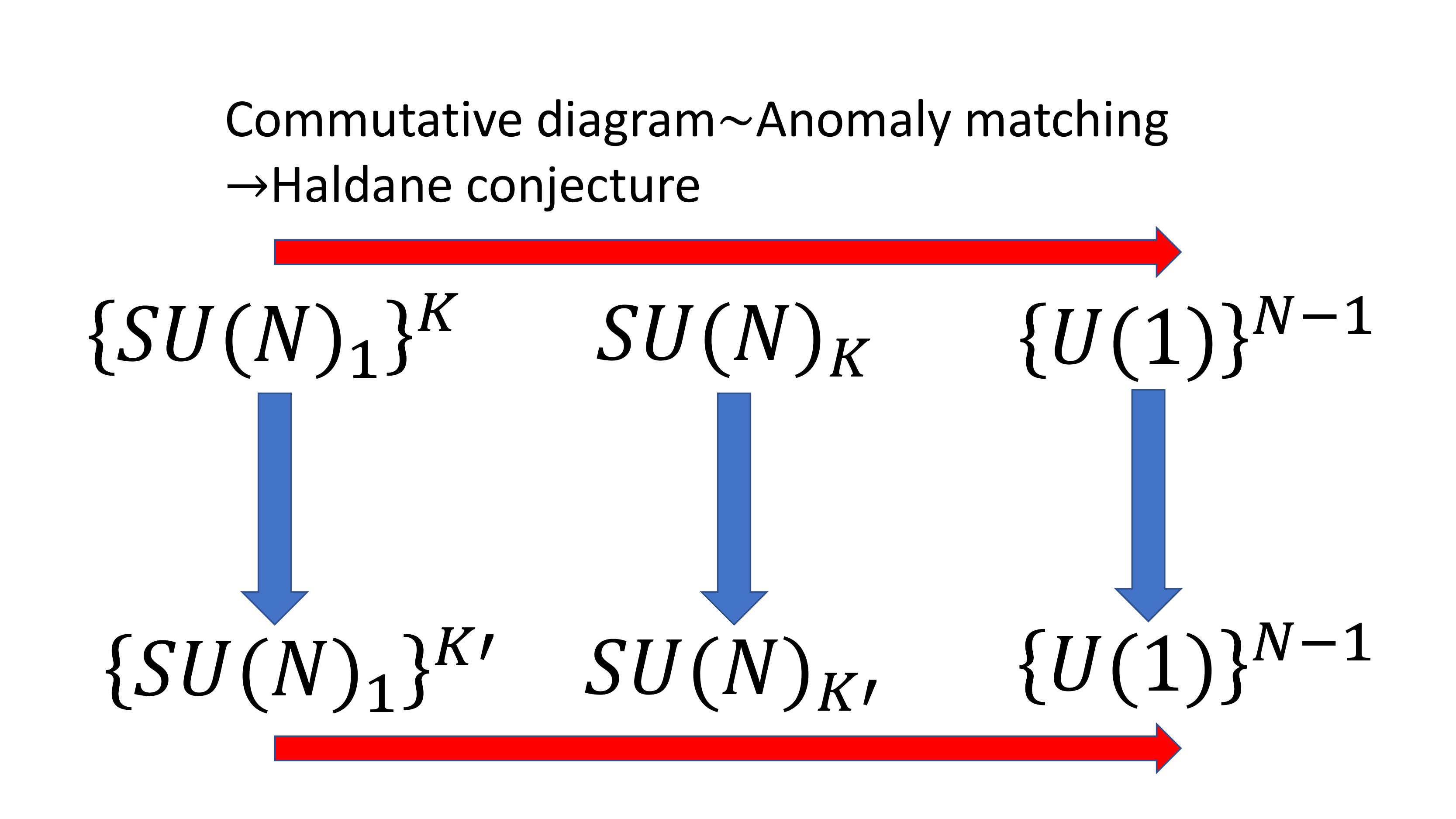}
\caption{RG flows of $SU(N)$ models. Assocativity of red arrows and blue arrows corresponds to the Haldane conjecture.}
\label{Haldane}
\end{center}
\end{figure}

Because of the $Z_{N}$ anomaly matching, one expects $K-K' (\text{mod}. N)$ periodicity naively. Especially, one can interprete the connectivity of $K=K' (\text{mod}N)$ as composite parafermion condensation of $\{ SU(N)_{1}\}^{K-K'}$ model.

The above analysis is coming from anomaly matching. Related discussion (and diagram) can be seen in several works, \cite{Cabra:1998vw,Lecheminant:2015iga,Cheng:2022sgb}, for example. The RG domain wall interpretation of the model, $\{ SU(N)_{1}\}^{K+K'}$ predicts another periodicity $K+K' \ (\text{mod}. N)$. Interestingly, this periodicity is different from the above anomaly matching argument.

Here we comment on the structure of coset representations in \cite{Fuji_2017}.
The mapping from $\{SU(N)_{1}\}^{K}$ to the following coset model is,

\begin{equation}
\{ SU(N)_{1} \}^{K} \sim \{ U(1) \}^{N-1}\times \frac{SU(N)_{K}}{\{U(1)\}^{N-1}}\times \frac{SU(K)_{N}}{\{U(1)\}^{K-1}}
\end{equation}
where $ \frac{SU(N)_{K}}{\{U(1)\}^{N-1}}$ and $\frac{SU(K)_{N}}{\{U(1)\}^{K-1}}$ are known as $Z_{K}$ Gepner parafermion and $Z_{N}$ Gepner parafermion respectively\cite{Gepner:1987sm}. It should be noted that one can expect $ \frac{SU(N)_{K}}{\{U(1)\}^{N-1}}$ as $Z_{N}$ anomaly free and $\frac{SU(K)_{N}}{\{U(1)\}^{K-1}}$ as $Z_{K}$ anomaly free as in the $SU(2)$ case\cite{Cabra:1998vw}. Interestingly, the Gepner parafermion part contains $Z_{K}$ simple current which is not evident from the original product theory. 

The RG flow of this model to $SU(N)_{K}$ can be expressed as,
\begin{equation}
\begin{split}
& \{ U(1) \}^{N-1}\times \frac{SU(N)_{K}}{\{U(1)\}^{N-1}}\times \frac{SU(K)_{N}}{\{U(1)\}^{K-1}}  \\
&\rightarrow \{ U(1) \}^{N-1}\times \frac{SU(N)_{K}}{\{U(1)\}^{N-1}}\sim SU(N)_{K}
\end{split}
\end{equation}
Hence the $Z_{N}$ anomaly of the model coming from $\frac{SU(K)_{N}}{\{U(1)\}^{K-1}}$ should be transfered to $\{U(1)\}^{N-1}$ part. It is also interesting to note that the level-rank duality of $SU(N)_{K}$ and $SU(K)_{N}$ explains the emergence of the $Z_{K}$ symmetry which was not evident from the original product theory $\{ SU(N)_{1}\}^{K}$. By gapping the parafermion part, one may obtain the $SU(N)$ Haldane conjecture. Unfortunately, this final step seems to require some complicated calculations. In principle, by generalizing the fermionization procedure in \cite{Alvarez-Gaume:1986nqf,Yao:2019bub}, or directly implementing the momentum shift property of LSM operator in $SU(N)_{1}\sim \{ U(1)\}^{N-1}$ CFT, one can represent the total picture in CFT determined by the anomaly. In this approach, one can obtain the table of the possible conformal dimensions of perturbations which depend on the generalized Luttinger parameter for $\{ U(1)\}^{N-1}$ CFT determined by the anisotropy parameters of the $SU(N)$ lattice model.

\begin{figure}[htbp]
\begin{center}
\includegraphics[width=0.5\textwidth]{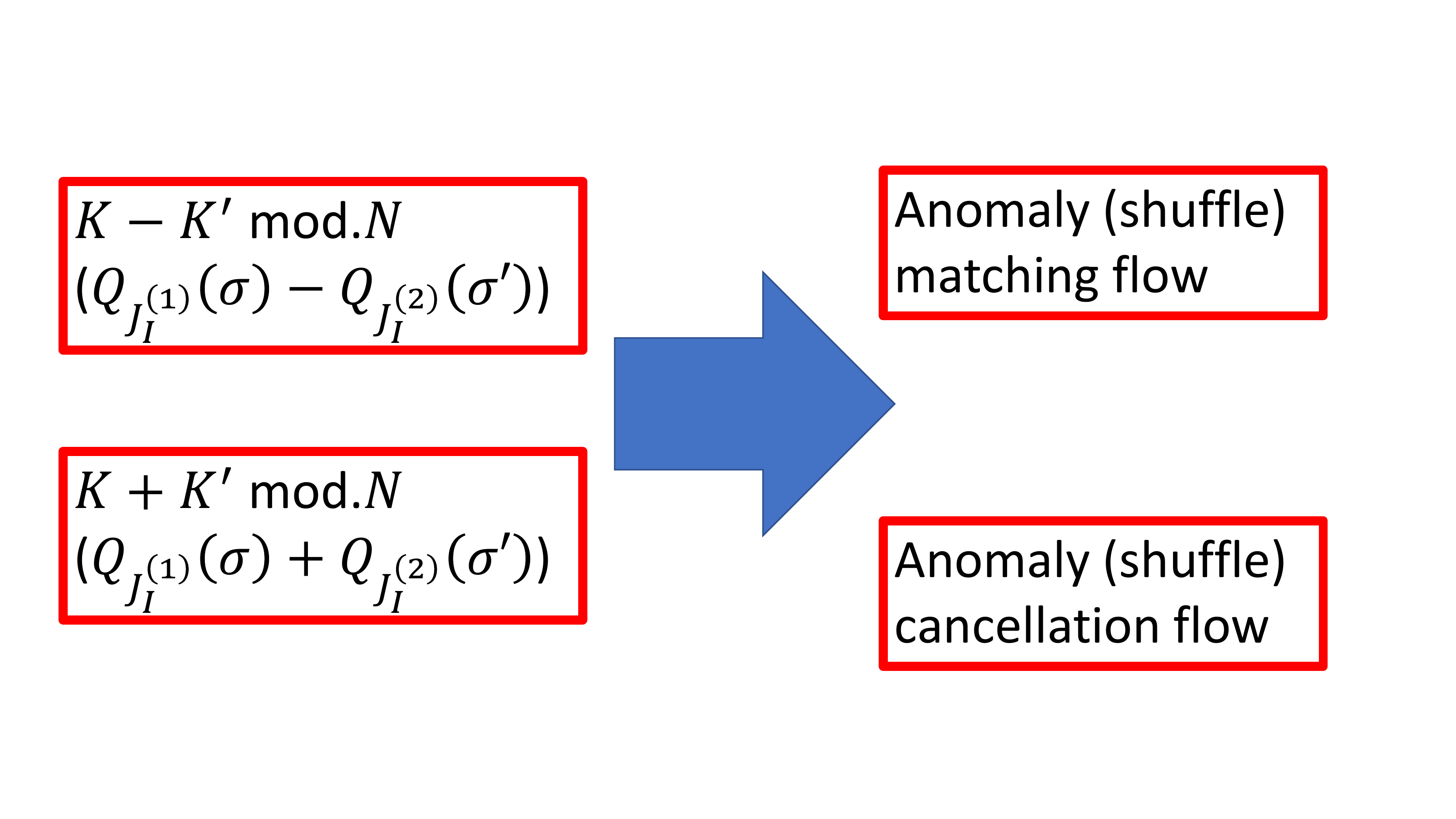}
\caption{Summary of matching and cancellation flows. Similar to anomaly analysis by considering the simple charges, one may be possible to analyze the Galois shuffle property by considering the monodromy charges.}
\label{Haldane}
\end{center}
\end{figure}

Consequently, there exists at least two periodicity $K-K' (\text{mod.}N)$ and $K+K' (\text{mod.}N)$ for $\{ SU(N)_{1}\}^{K}$ model\footnote{By considering the level-rank duality of $SU(N)_{K}$, one can obtain $N+N' (\text{mod.}K)$ and $N-N' (\text{mod.}K)$ periodicity of $SU(N)_{K}$ model by fixing $K$. We thank Yuan Yao for the helpful comments on this aspect.}. These two periodicities come from slightly different mechanisms, and we call the corresponding integrable flow for $K-K'=0 (\text{mod.}N)$ and $K+K'=0 (\text{mod.}N)$  as "anomaly matching flow" and "anomaly cancellation flow " respectively.

As we have discussed in section 3, one can consider similar matching and cancellation for Galois shuffle in RG connective theories. Let us introduce the two theory $M_{1}$ with simple currents $J_{I}^{(1)}$ and shuffle generator $\sigma$, and $M_{2}$ with simple current $J_{I}^{(2)}$ and the shuffle generator $\sigma'$. Then, one can express the (hopefully integrable) flow with $Q_{J_{I}^{(1)}}(\sigma)-Q_{J_{I}^{(2)}}(\sigma')=0$ as "shuffle matching flow" and that with $Q_{J_{I}^{(1)}}(\sigma)+Q_{J_{I}^{(2)}}(\sigma')=0$ as "shuffle cancellation flow". Hopefully, this may give a new criterion to establish a classification of nonunitary theories.

In summary, we have discussed the importance of the RG analysis of CFT described by the products of simple theories possibly with large central charge, with emphasis on anomaly\footnote{In this sense, studying CFTs with large central charge or gigantic symmetries are reasonable.}. The essence of our proposal is simple in that RG started from sufficient bulk and boundary degrees of freedom should explain a class of nontrivial symmetry-related phenomena, such as $SU(N)$ Haldane conjecture.

\section{Conclusion}
\label{conclusion}
In this work, we have constructed topologically ordered systems with simple current anomaly free theory. The cylinder partition functions constructed in this work satisfy the necessary and sufficient conditions for the stability in the existing literature almost automatically. It may be worth stressing that we have never used the category theory at least directly except for assuming the bosonic charge conjugated partition functions. We hope this work may give a new bridge to connect experimentalists in condensed matter physics and theorists in all related areas such as high energy physics and mathematical physics.  

We have viewed several modern aspects of the topological phase, such as confinement, fractional supersymmetry, and non-locality, with emphasis on the quantum anomaly in the simple models constructed by the bulk-edge correspondence. We assume $CFT_{D}/BQFT_{D+1}$ correspondence known as bulk-edge correspondence and its RG flow to $CFT_{D}/TQFT_{D+1}$ correspondence. This proposition is implicitly assumed by a lot of existing literature, typically in the name of $AdS/CFT$ or $AdS/CMT$ correspondence.
So far, because we have constructed everything almost automatically, there appears a simple question. We have assumed the protectedness of edge modes, but it is not clear whether this can be shown by using a traditional RG argument, in the language of bulk and boundary RG flow. Hence for the future problem, as the author has emphasized in the related work\cite{Fukusumi_2022_f}, it may become important to check the extent of the protectedness by using field theory and numerical simulation of lattice models. For example, developments of the truncated conformal space approach (TCSA) in higher dimensional CFT with boundary seem necessary\cite{James_2018}. However, the work for higher dimensional CFT is limited\cite{Hogervorst:2014rta}. Moreover, there exist some open problems for the appropriate definition of the operators at the boundary even in lower dimensions. For example, one cannot use the Kac identification or other useful property of the bulk local operators\cite{PhysRevLett.119.191601}. This type of difficulty has never captured sufficient attention in the fields, but to study edge modes of topological phase by using RG, it seems impossible to avoid this difficulty.

As a concluding remark, we emphasize the analogy with FQHE and string theory, known as 2d quantum gravity coupled to matter. In FQHE, the most essential part of the construction is the single-valuedness of the wavefunction. This restricts the building structure of CFT correlation functions as a wavefunction and results in the nontrivial emergence of fractional flux quantum. Historically, the single-value condition appeared in the study of strong interaction and non QFT (or non-Lagrangian) aspects appear naturally. As we have discussed in the series of previous works\cite{Fukusumi_2022,Fukusumi_2022_f}, there exist common properties between FQHE and 2d quantum gravity coupled to matter\cite{Polyakov:1981rd}. We summarize the point in the following figure (FIG. \ref{KPZ}). The appearance of fractional flux quantum in FQHE system corresponds to the appearance of gravitational dressing by the KPZ relation\cite{David:1988hj,Knizhnik:1988ak,Distler:1988jt}. The correspondence between the fusion rules in the original matter CFT and 2d gravity coupled to matter is called ground ring \cite{Lian:1991ju,Lian:1991gk,Kutasov:1991qx}, and this corresponds to the abelian property of the multicomponent bosons in FQHE. There exist related works which point out connections between FQHE and string theories by using the holography principle \cite{Fujita:2009kw,Ryu:2010fe,Bayntun:2010nx,Karch:2010mn,Melnikov:2012tb,Wu:2014dha,Mezzalira:2015vzn}. One can say that our argument is a more direct approach to construct the corresponding systems in operator formalism in quantum method whereas these works use the Lagrangian formalism which corresponds to higher dimensional classical methods.

\begin{figure}[htbp]
\begin{center}
\includegraphics[width=0.5\textwidth]{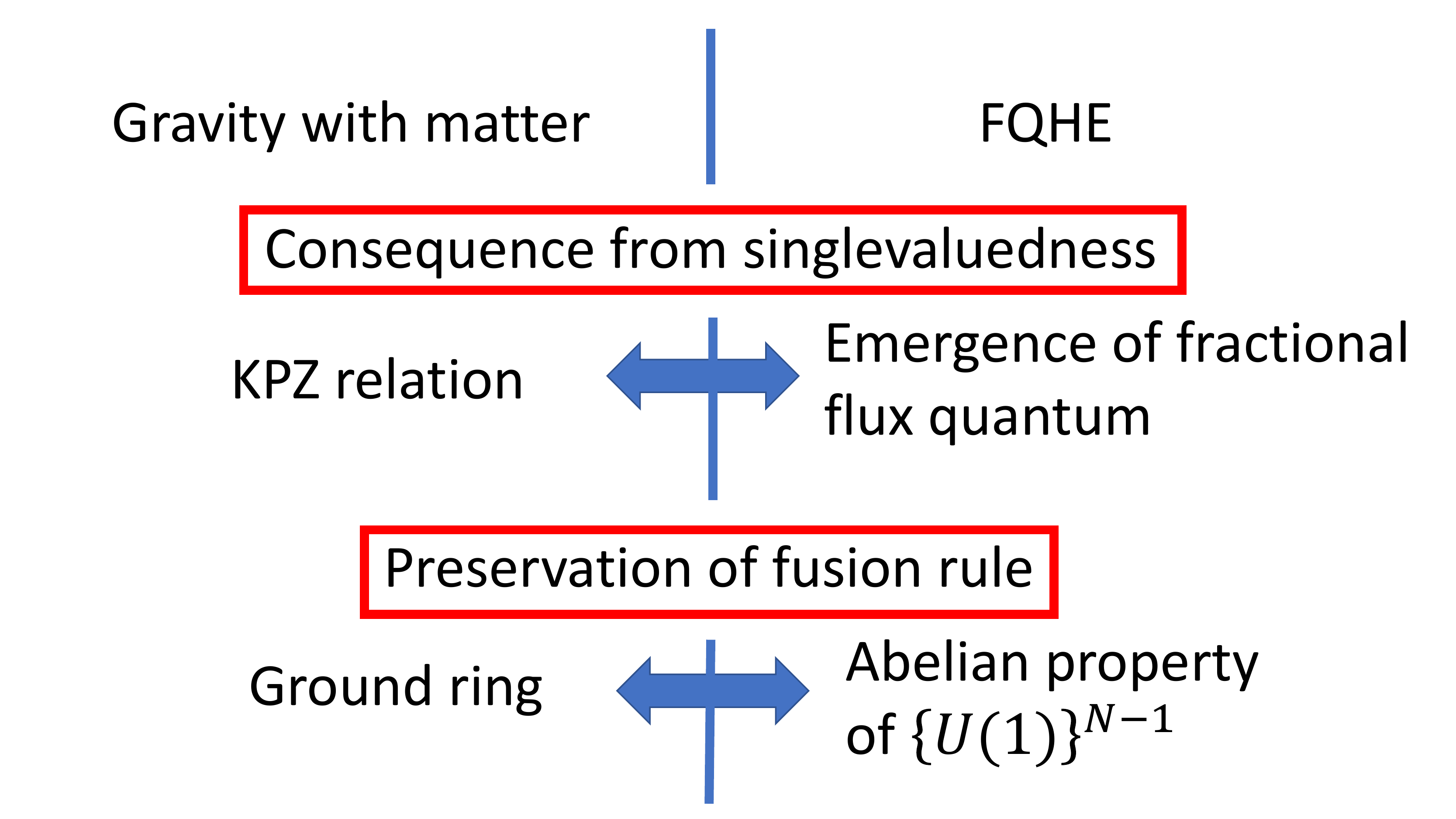}
\caption{Analogy between 2d quantum gravity couple to matter and FQHE. FQHE can be considered as a gauge theory analogous to string theory, in the context of gauge/gravity duality.}
\label{KPZ}
\end{center}
\end{figure}

In statistical mechanics, 2d quantum gravity coupled to matter is interpreted as coupling of matter and random surface\cite{Watabiki:1993fk}. Hence one can interpret gravity as random objects, which can be realized in a probabilitic model. For example, one can find several classes of statistical mechanical models corresponding to this gravity, such as whole-plane or backward Schramm-Loewner evolution\cite{https://doi.org/10.48550/arxiv.1211.2451,Katori_2020}, Laplacian growth\cite{Alekseev:2017nwo}, log-correlated random energy model\cite{Cao:2016hvd}, and Gaussian multiplicative chaos\cite{https://doi.org/10.48550/arxiv.1305.6221}. Hence, if there exists a similar interpretation for the 2+1 dimensional system constructed from 2d quantum gravity coupled to matter as for FQHE, one can expect applications of such models to condensed matter systems with impurities or random potential for example. Similar approaches can be seen in the random Dirac fermion model related to localization problem \cite{Kogan:1996wk, Bhaseen:2000mi, Bhaseen:2000gr}. Moreover, at first sight, this coupling to gravity automatically cancels the anomaly of simple currents even when the original matter CFT is anomalous. Hence further studies of 2d quantum gravity constructed from a composite parafermionic theory may give a unified understanding and construction of general topological ordered systems.

\section{Acknowlegement}
The author thanks Yuji Tachikawa and Yunqin Zheng for the fruitful collaboration in the previous project. He also thanks Yang Bo for the discussion and collaboration with the other related projects.
 He thanks Yuan Yao for the stimulating discussions about parafermionization and Yohei Fuji for introducing the basic aspects of $Z_{N}$ fractional quantum Hall effects, and Shunsuke Furukawa for the discussion about Renyification. He also thanks Ken Kikuchi for the discussions and his comments on the manuscript. 

\appendix

\section{Summary of parafermionic partition function}

In this section, we summarize the partition functions which we have introduced in the main text. First, let us introduce the $Z_{N}$ model with charge conjugated modular invariant,
\begin{equation}
Z=\sum_{i,p}\chi_{i,p}\overline{\chi}_{i,-p}+\sum_{a}|\chi_{a}|^{2}
\end{equation}
where $\{ i,p\}_{p=0}^{N-1}$ is the label of the $Z_{N}$ noninvariant states, and $\{ a\}$ is that of the $Z_{N}$ invariant states and $\chi$ ($\overline{\chi}$) is the corresponding chiral (antichiral) character with the modular parameter $\tau$ ($\overline{\tau}$). 

In the anomaly free case, the composite parafermionic partition function for each simple charge sector $\{ Q_{I} \}_{I=1}^{N-1}$ where $\{ Q_{I} \}_{I=0}^{N-1}$ corresponds to the simple charge $\{ Q_{J_{I}}(\alpha)\}_{I=1}^{N-1}$ for the operator labeled $\alpha$ under $Z_{N}$ simple current $\{ J_{I}\}_{I=1}^{N-1}$ is expressed as,
\begin{equation}
Z_{\{Q_{I}\}}=\sum_{i \in\{Q_{J_{I}}(i)=Q_{I}\}}|\sum_{p}\chi_{i,p}|^{2}+N\sum_{a\in\{Q_{J_{I}}(a)=Q_{I}\}}|\chi_{a}|^{2}.
\end{equation}
By considering the fractional flux quantum insertion corresponding to above $\{ Q_{I}\}_{I=1}^{N-1}$ and assigning modular $T$ or $T^{2}$ invariance, one can obtain the $Z_{N}$ FQHE partition functions in the cylinder geometry in the main section. 
The total (modular noninvariant) partition function is 
\begin{equation}
Z=\sum_{\{Q_{I}\}}\left(\sum_{\{r_{I}\}}\left(\sum_{i}|\sum_{p}\Xi_{i,p,\{ r_{I} \}}|^{2}+N\sum_{a}|\Xi_{a,\{ r_{I}\}}|^{2}\right)\right)
\end{equation}
where we have taken the same notations in the main text in section \ref{FQHE}, and introduced a character $\Xi_{i,p,\{ r_{I}\}}= \sum_{p'}\Xi_{i,p,\{r_{I}\}}^{p'}$.

In an anomalous case, the chiral parafermionization gives the following partition function,
\begin{equation}
Z_{\{Q_{I}\}}
=\sum_{i, p\in\{Q_{J_{I}}(i,p)=Q_{I}\}}\left(\chi_{i,p}\right)\left(\sum_{p'}\overline{\chi}_{i,p'}\right),
\end{equation}
by replacing $\tau\leftrightarrow \overline{\tau}$, one can obtain the partition function corresponding to antichiral parafermionization.

Here, we comment on two straightforward generalizations of our approach. First, by replacing simple charge with monodromy charge, one can easily obtain the parafermionic partition function for nonunitary theories. Second, by considering operator counting of the bosonic model, one can obtain a similar form of parafermionization in general.

\section{Comments on subalgebra structure in fusion rule}

As we have shown, the introduction of simple charges of the fields naturally extracts the subalgebra structure of the fusion rule of CFT. In our construction, there exists the other fundamental structure, parafermionic parity. For simplicity, let us focus on the $Z_{2}$ case. In this fermionic theory, the fermionic parity of an operator is even or odd. Hence assuming that this parity is well-defined, one can obtain the set of the parity even fields as a subalgebra, because it satisfies,
\begin{equation} 
\phi_{\alpha}^{\text{even}}\times \phi_{\alpha'}^{\text{even}}=\sum_{\beta}N_{\alpha,\alpha'}^{\beta}\phi_{\beta}^{\text{even}}
\end{equation}

As one can obtain untwisted part of the fusion rule by the parafermionization, it might be interesting to consider whether it is possible to obtain this subalgebra by considering the generalized gauging\footnote{In finalizing this work, we have noticed related literature considering the similar (or possibly the same) operation \cite{Runkel:2022fzi}.}. To distinguish parity odd and even states, it might be useful to consider parity operators $(-1)^{F}$, $(-1)^{\overline{F}}$,  $(-1)^{F+\overline{F}}$ where $F$ (or $\overline{F}$) is chiral (or antichiral) fermionic parity operator. As a simple example, we note the appearance of Fibonacci anyon, with fusion rule $\tau\times \tau =I+\tau$, from the parafermionic parity 0 sector of the three state Potts model and from the untwisted fermionic parity 0 sector of the tricritical Ising model. One can systematically obtain such subalgebra structures by considering the simple charge and the parafermionic parity of the model.

In the existing literature, there exist several proposals to insert topological symmetry operators, but we comment here on other more traditional interpretation based on Hamiltonian formalism. As the author has discussed in \cite{Fukusumi:2020irh}, the insertion of the topological symmetry operator can be understood as the insertion of generalized twisted boundary condition or integral of motion (We denote the topological symmetry operators as $Q_{s}$ and $Q_{t}$ corresponding to each direction). This interpretation can be seen in the study of Temperley-Lieb algebra description of lattice models and the corresponding CFTs\cite{https://doi.org/10.48550/arxiv.2003.11293}. It should be noted, these operations themselves cannot be treated in a local manner in quantum systems\cite{Roy:2021jus,Roy:2021xul}. Whereas this insertion of topological symmetry operator in the Lagargian formalism is established, there exists another "gauging" operation like in lattice gauge field theory,
\begin{equation} 
H'=H_{CFT}+gQ_{t}+\text{h.c.c.}. 
\label{lattice_gauging}
\end{equation}

By taking the large $|g|$ limit, one can obtain topological sectors of CFT corresponding to the eigenvalue of the integral of motions $Q_{t}$. This can be interpreted as a generalization of Gauss law in the lattice gauge models. One can easily generalize this analysis to higher-form objects in general dimensions. It might be interesting whether this "gauging" can reproduce other types of local models, such as anyonic models\cite{Feiguin:2006ydp}. 

\subsection{Ising chain and gauging}
As a simplest example for the gauging Eq. \eqref{lattice_gauging}, let us consider the transverse Ising model at criticality and the corresponding partition function by CFT.
The lattice Hamiltonian of the model is,
\begin{equation}
H_{\text{Ising}}=-\sum_{l}\left( \sigma_{l}\sigma_{l+1}+\tau_{l}\right).
\end{equation} 

The $Z_{2}$ symmetry operation which corresponds to topological defect $Q_{\epsilon}$ for Ising CFT is,
\begin{equation}
Q_{\epsilon}=\prod_{l}\tau_{l},
\end{equation}

This commute with Hamiltonian and one can consider the spectrum of each charge sector with eigenvalue $\pm 1$.
The partition function of the original Hamiltonian with periodic boundary condition or twisted boundary condition is,
\begin{align}
Z_{\ ,+}&= |\chi_{0}|^{2}+|\chi_{\epsilon}|^{2}+|\chi_{\sigma}|^{2},\\
Z_{\ ,-}&=\chi_{0}\overline{\chi_{\epsilon}}+\chi_{\epsilon}\overline{\chi_{0}}+|\chi_{\sigma}|^{2},
\end{align}
It should be noted that each sector can be labeled by the eigenvalue of $Q_{\epsilon}$ or simple charge in the main section. Hence the partition function can be decomposed as,

\begin{align}
Z_{+ ,+}&= |\chi_{0}|^{2}+|\chi_{\epsilon}|^{2},\\
Z_{+ ,+}&=|\chi_{\sigma}|^{2},\\
Z_{+ ,-}&=\chi_{0}\overline{\chi_{\epsilon}}+\chi_{\epsilon}\overline{\chi_{0}}, \\
Z_{- ,-}&=|\chi_{\sigma}|^{2},
\end{align}
where $Z_{\ ,\pm}=Z_{+,\pm}+Z_{-, \pm}$.

Here we introduce the "gauged" Hamiltonian,

\begin{equation}
H_{\text{Ising}}+gQ_{\epsilon}
\end{equation}

By taking $|g|$ large limit, splitting by simple charge becomes large and the partition function $Z_{\pm, \pm}$ and $Z_{\pm, \mp}$ appear. There exist following remarkable relations,
\begin{align}
Z_{NS}&=Z_{+,+}+Z_{+,-}=|\chi_{0}+\chi_{\epsilon}|^{2}, \\
Z_{R}&= Z_{-,+}+Z_{-,-}=2|\chi_{\sigma}|^{2}.
\end{align}
The summation $Z_{\pm, +}+Z_{\pm, -}$ was taken corresponding to the Gauss law. This is analogous to the various models by considering $Z_{2}$ gauging, and one can see the nonlocal aspects of the gauged theory as an addition of $Q_{\epsilon}$ started from a conformal field theory. Moreover, as we have discussed in the main text and the related paper by the author\cite{Fukusumi_2022_f}, the energy splitting by $g$ and duality (or supersymmetry) between NS and R sector is analogous to quark confinement. In this section, we have mainly used the operator formalism, but this appearance of Gauss law in the model can be considered as a Hamiltonian and discrete symmetry analog of the established proposals of chiral gauging in the Lagrangian formalism\cite{Faddeev:1984jp,Hosono:1987na,Fujiwara:1988zc} 

By considering the $Z_{N}$ symmetry operator and the lattice realization of composite parafermions, one can easily generalize the discussion above.

\section{Composite parafermion on the lattice}
In analogy with CFT, one can construct lattice composite parafermion by coupling the lattice parafermion $N$ times in general $N$ and $N/2$ times for $N$ even with the same chirality. First, let us consider the general case, and assume $Z_{N}$ parafermionic lattice systems $\{\gamma_{i}\}$ with $NL$ site. Then, one can define the following lattice composite parafermion
\begin{equation}
\Gamma_{i}=\gamma_{i}\times \gamma_{i+L}\times ... \times \gamma_{i+(N-1)L}=\prod_{j=0}^{N-1}\gamma_{i+jL}
\end{equation}
where $i$ takes $i=1$ to $i=L$.
This is a lattice analog of the composite parafermion with the following algebraic relation,
\begin{align}
\Gamma_{i}\Gamma_{j}=\Gamma_{j}\Gamma_{i}, \\
\Gamma_{i}^{N}=1, \Gamma_{i}^{\dagger}=\Gamma_{i}^{-1},
\end{align} 

For $N$ even case, by coupling the parafermion $N/2$ times, one can obtain the similar relations,
\begin{align}
\Gamma_{i}\Gamma_{j}+\Gamma_{j}\Gamma_{i}=0, \\
\Gamma_{i}^{N}=1, \Gamma_{i}^{\dagger}=\Gamma_{i}^{-1}.
\end{align} 

For simplicity, let us observe the nonuniqueness of composite parafermionization for $N=2$ and study its relation to $SU(2)_{K}$ Haldane conjecture. In $SU(2)_{K}$ model, there exists periodicty $K(\text{mod}.4)$ for statistcs of the $Z_{2}$ simple current, and for $K=0 (\text{mod}.2)$ corresponds to anomaly free case. Hence we introduce $K=2S$ where $S$ is an integer and corresponds to the spin of the Heisenberg spin chain. In this setting there exists another periodicity $S (\text{mod}2)$ related to hidden $Z_{2}\times Z_{2}$ symmetry\cite{MOshikawa_1992}. Then, one can interpret $S$ as the number of Majorana fermions because of anomaly matching, this corresponds to the above two different types of composite fermionization. As we have stressed in the main text, the nontrivial hidden symmetries can be more evident by considering the theory with a larger central charge.

For a future problem, it might be interesting to study the phase diagram of systems which is described by such composite particles. For example, our formulation may give an understanding of the lattice fractional supersymmetric model in \cite{Mong:2014ova}. We expect the appearance of fractional spin particles by twisted boundary conditions, which are commonly used in the bosonic Coulomb gas representation, as we have discussed in the main text.

\bibliography{parafermion}

\end{document}